\documentstyle [12pt,amsfonts] {article}
\input epsf

\newcommand{\beq}{\begin{equation}}
\newcommand{\eeq}{\end{equation}}
\newcommand{\beqs}{\begin{eqnarray}}
\newcommand{\eeqs}{\end{eqnarray}}

\catcode`@=11
\@addtoreset{equation}{section}
\@addtoreset{equation}{subsection}
\def\theequation{\ifnum\value{section}=0 \arabic{equation}\ignorespaces
\else \ifnum\value{section}=-1 A.\arabic{equation}\ignorespaces
\else \ifnum\value{subsection}=0 \thesection.\arabic{equation}\ignorespaces
\else \thesection.\arabic{subsection}.\arabic{equation}\ignorespaces
                           \fi
                      \fi
                 \fi}
\catcode`@=12

\begin{document}

\def\thefootnote{\fnsymbol{footnote}}

\baselineskip 6.0mm

\vspace{4mm}

\begin{center}

{\Large \bf Ground State Entropy of the Potts Antiferromagnet on Triangular 
Lattice Strips}

\vspace{8mm}

\setcounter{footnote}{0}
Shu-Chiuan Chang\footnote{email: shu-chiuan.chang@sunysb.edu} and 
\setcounter{footnote}{6}
Robert Shrock\footnote{email: robert.shrock@sunysb.edu}

\vspace{6mm}

C. N. Yang Institute for Theoretical Physics  \\
State University of New York       \\
Stony Brook, N. Y. 11794-3840  \\

\vspace{10mm}

{\bf Abstract}
\end{center}

We present exact calculations of the zero-temperature partition function
(chromatic polynomial) $P$ for the $q$-state Potts antiferromagnet on
triangular lattice strips of arbitrarily great length $L_x$ vertices and of
width $L_y=3$ vertices and, in the $L_x \to \infty$ limit, the exponent of the
ground-state entropy, $W=e^{S_0/k_B}$.  The strips considered, with their
boundary conditions ($BC$) are (a) $(FBC_y,PBC_x)=$ cyclic, (b)
$(FBC_y,TPBC_x)=$ M\"obius, (c) $(PBC_y,PBC_x)=$ toroidal, and (d)
$(PBC_y,TPBC_x)=$ Klein bottle, where $F$, $P$, and $TP$ denote free, periodic,
and twisted periodic.  Exact calculations of $P$ and $W$ are also given for
wider strips, including (e) cyclic, $L_y=4$, and (f) $(PBC_y,FBC_x)=$
cylindrical, $L_y=5,6$. Several interesting features are found, including the
presence of terms in $P$ proportional to $\cos(2\pi L_x/3)$ for case (c).
The continuous locus of points ${\cal B}$ where $W$ is nonanalytic in the $q$
plane is discussed for each case and a comparative discussion is given of the
respective loci ${\cal B}$ for families with different boundary conditions.
Numerical values of $W$ are given for infinite-length strips of various widths
and are shown to approach values for the 2D lattice rapidly.  A remark is also
made concerning a zero-free region for chromatic zeros. 

\vspace{16mm}

\pagestyle{empty}
\newpage

\pagestyle{plain}
\pagenumbering{arabic}
\renewcommand{\thefootnote}{\arabic{footnote}}
\setcounter{footnote}{0}

\section{Introduction} 

The $q$-state Potts antiferromagnet (AF) \cite{potts,wurev} exhibits nonzero
ground state entropy, $S_0 > 0$ (without frustration) for sufficiently large
$q$ on a given lattice $\Lambda$ or, more generally, on a graph $G$.  This is
equivalent to a ground state degeneracy per site $W > 1$, since $S_0 = k_B \ln
W$.  Such nonzero ground state entropy is important as an exception to the
third law of thermodynamics.  There is a close connection with graph theory
here, since the zero-temperature partition function of the above-mentioned
$q$-state Potts antiferromagnet on a graph $G$ satisfies
\beq
Z(G,q,T=0)_{PAF}=P(G,q)
\label{zp}
\eeq
where $P(G,q)$ is the chromatic polynomial expressing the number of ways of
coloring the vertices of the graph $G$ with $q$ colors such that no two
adjacent vertices have the same color (for reviews, see
\cite{rrev}-\cite{bbook}).  The minimum number of colors necessary for such a
coloring of $G$ is called the chromatic number, $\chi(G)$. 
Thus\footnote{\footnotesize{At certain special
points $q_s$ (typically $q_s=0,1,.., \chi(G)$), one has the noncommutativity of
limits $\lim_{q \to q_s} \lim_{n \to \infty} P(G,q)^{1/n} \ne \lim_{n \to
\infty} \lim_{q \to q_s}P(G,q)^{1/n}$, and hence it is necessary to specify the
order of the limits in the definition of $W(\{G\},q_s)$ \cite{w}.  We use the
first order of limits here; this has the advantage of removing certain isolated
discontinuities in $W$.}}
\beq
W(\{G\},q) = \lim_{n \to \infty} P(G,q)^{1/n}
\label{w}
\eeq 
where $n=v(G)$ is the number of vertices of $G$ and $\{G\} = \lim_{n \to
\infty}G$.  Since $P(G,q)$ is a polynomial, one can generalize $q$ from
${\mathbb Z}_+$ to ${\mathbb C}$.  The zeros of $P(G,q)$ in the complex $q$
plane are called chromatic zeros; a subset of these may form an accumulation
set in the $n \to \infty$ limit, denoted ${\cal B}$, which is the continuous
locus of points where $W(\{G\},q)$ is
nonanalytic.~\footnote{\footnotesize{Although it does not happen in the cases
considered here, for some families of graphs ${\cal B}$ may be null, and $W$
may also be nonanalytic at certain discrete points.}}  The maximal region in
the complex $q$ plane to which one can analytically continue the function
$W(\{G\},q)$ from physical values where there is nonzero ground state entropy
is denoted $R_1$.  The maximal value of $q$ where ${\cal B}$ intersects the
(positive) real axis is labelled $q_c(\{G\})$.  This point is important since
it separates the interval $q > q_c(\{G\})$ on the positive real $q$
axis where the Potts model (with $q$ extended from ${\mathbb Z}_+$ to ${\mathbb
R}$) exhibits nonzero ground state entropy (which increases with $q$,
asymptotically approaching $S_0 = k_B \ln q$ for large $q$, and which for a
regular lattice $\Lambda$ can be calculated approximately via large--$q$ series
expansions) from the interval $0 \le q \le q_c(\{G\})$ in which $S_0$ has a
different analytic form.

\unitlength 1.3mm
\begin{picture}(100,20)
\multiput(0,0)(10,0){5}{\circle*{2}}
\multiput(0,10)(10,0){5}{\circle*{2}}
\multiput(0,20)(10,0){5}{\circle*{2}}
\multiput(0,0)(10,0){5}{\line(0,1){20}}
\multiput(0,0)(0,10){3}{\line(1,0){40}}
\put(0,10){\line(1,1){10}}
\multiput(0,0)(10,0){3}{\line(1,1){20}}
\put(30,0){\line(1,1){10}}
\put(-2,-2){\makebox(0,0){9}}
\put(8,-2){\makebox(0,0){10}}
\put(18,-2){\makebox(0,0){11}}
\put(28,-2){\makebox(0,0){12}}
\put(38,-2){\makebox(0,0){9}}
\put(-2,12){\makebox(0,0){5}}
\put(8,12){\makebox(0,0){6}}
\put(18,12){\makebox(0,0){7}}
\put(28,12){\makebox(0,0){8}}
\put(38,12){\makebox(0,0){5}}
\put(-2,22){\makebox(0,0){1}}
\put(8,22){\makebox(0,0){2}}
\put(18,22){\makebox(0,0){3}}
\put(28,22){\makebox(0,0){4}}
\put(38,22){\makebox(0,0){1}}
\put(20,-8){\makebox(0,0){(a)}}

\multiput(60,0)(10,0){5}{\circle*{2}}
\multiput(60,10)(10,0){5}{\circle*{2}}
\multiput(60,20)(10,0){5}{\circle*{2}}
\multiput(60,0)(10,0){5}{\line(0,1){20}}
\multiput(60,0)(0,10){3}{\line(1,0){40}}
\put(60,10){\line(1,1){10}}
\multiput(60,0)(10,0){3}{\line(1,1){20}}
\put(90,0){\line(1,1){10}}
\put(58,-2){\makebox(0,0){5}}
\put(68,-2){\makebox(0,0){6}}
\put(78,-2){\makebox(0,0){7}}
\put(88,-2){\makebox(0,0){8}}
\put(98,-2){\makebox(0,0){1}}
\put(58,12){\makebox(0,0){9}}
\put(68,12){\makebox(0,0){10}}
\put(78,12){\makebox(0,0){11}}
\put(88,12){\makebox(0,0){12}}
\put(98,12){\makebox(0,0){9}}
\put(58,22){\makebox(0,0){1}}
\put(68,22){\makebox(0,0){2}}
\put(78,22){\makebox(0,0){3}}
\put(88,22){\makebox(0,0){4}}
\put(98,22){\makebox(0,0){5}}
\put(80,-8){\makebox(0,0){(b)}}
\end{picture}
\vspace*{3cm}

\begin{picture}(100,30)
\multiput(0,0)(10,0){5}{\circle*{2}}
\multiput(0,10)(10,0){5}{\circle*{2}}
\multiput(0,20)(10,0){5}{\circle*{2}}
\multiput(0,30)(10,0){5}{\circle*{2}}
\multiput(0,0)(10,0){5}{\line(0,1){30}}
\multiput(0,0)(0,10){4}{\line(1,0){40}}
\put(0,20){\line(1,1){10}}
\put(0,10){\line(1,1){20}}
\multiput(0,0)(10,0){2}{\line(1,1){30}}
\put(20,0){\line(1,1){20}}
\put(30,0){\line(1,1){10}}
\put(-2,-2){\makebox(0,0){1}}
\put(8,-2){\makebox(0,0){2}}
\put(18,-2){\makebox(0,0){3}}
\put(28,-2){\makebox(0,0){4}}
\put(38,-2){\makebox(0,0){1}}
\put(-2,12){\makebox(0,0){9}}
\put(8,12){\makebox(0,0){10}}
\put(18,12){\makebox(0,0){11}}
\put(28,12){\makebox(0,0){12}}
\put(38,12){\makebox(0,0){9}}
\put(-2,22){\makebox(0,0){5}}
\put(8,22){\makebox(0,0){6}}
\put(18,22){\makebox(0,0){7}}
\put(28,22){\makebox(0,0){8}}
\put(38,22){\makebox(0,0){5}}
\put(-2,32){\makebox(0,0){1}}
\put(8,32){\makebox(0,0){2}}
\put(18,32){\makebox(0,0){3}}
\put(28,32){\makebox(0,0){4}}
\put(38,32){\makebox(0,0){1}}
\put(20,-8){\makebox(0,0){(c)}}

\multiput(60,0)(10,0){5}{\circle*{2}}
\multiput(60,10)(10,0){5}{\circle*{2}}
\multiput(60,20)(10,0){5}{\circle*{2}}
\multiput(60,30)(10,0){5}{\circle*{2}}
\multiput(60,0)(10,0){5}{\line(0,1){30}}
\multiput(60,0)(0,10){4}{\line(1,0){40}}
\put(60,20){\line(1,1){10}}
\put(60,10){\line(1,1){20}}
\multiput(60,0)(10,0){2}{\line(1,1){30}}
\put(80,0){\line(1,1){20}}
\put(90,0){\line(1,1){10}}
\put(58,-2){\makebox(0,0){1}}
\put(68,-2){\makebox(0,0){2}}
\put(78,-2){\makebox(0,0){3}}
\put(88,-2){\makebox(0,0){4}}
\put(98,-2){\makebox(0,0){5}}
\put(58,12){\makebox(0,0){5}}
\put(68,12){\makebox(0,0){6}}
\put(78,12){\makebox(0,0){7}}
\put(88,12){\makebox(0,0){8}}
\put(98,12){\makebox(0,0){1}}
\put(58,22){\makebox(0,0){9}}
\put(68,22){\makebox(0,0){10}}
\put(78,22){\makebox(0,0){11}}
\put(88,22){\makebox(0,0){12}}
\put(98,22){\makebox(0,0){9}}
\put(58,32){\makebox(0,0){1}}
\put(68,32){\makebox(0,0){2}}
\put(78,32){\makebox(0,0){3}}
\put(88,32){\makebox(0,0){4}}
\put(98,32){\makebox(0,0){5}}
\put(80,-8){\makebox(0,0){(d)}}
\end{picture}
\vspace*{3cm}

\begin{figure}[h]
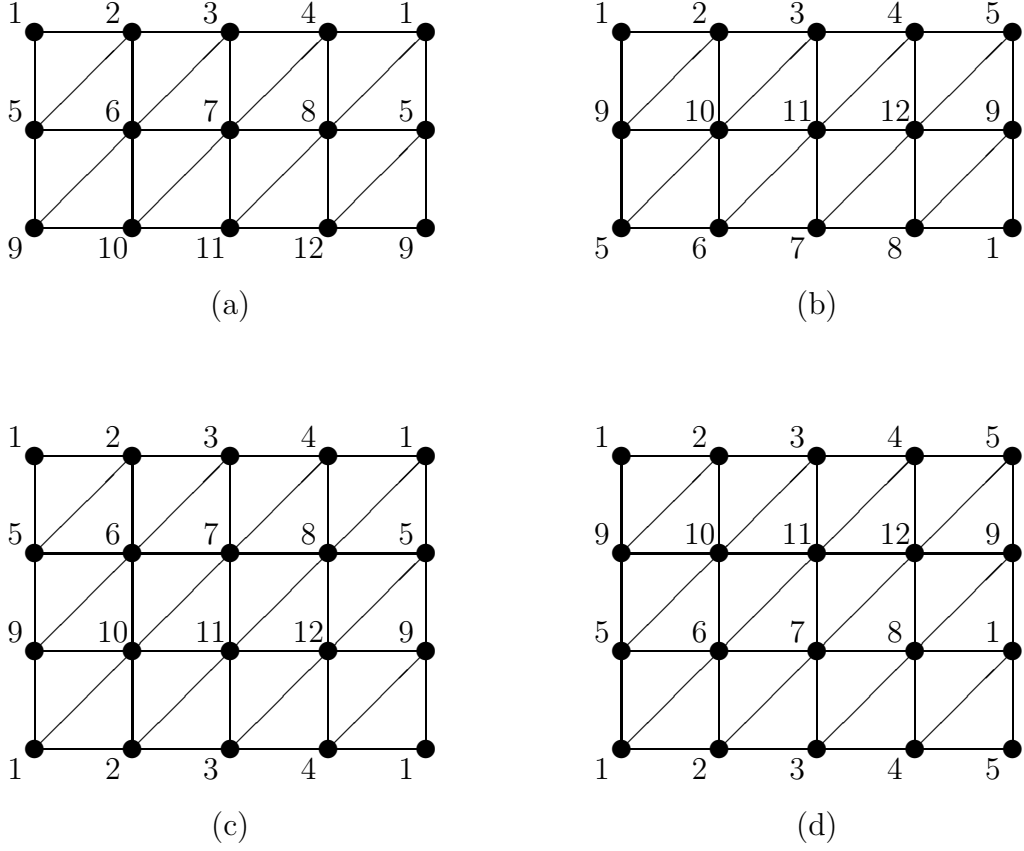

\caption{\footnotesize{Illustrative strip graphs of the triangular lattice with
width $L_y=3$ and length $L_x=4$ having the following
boundary conditions: (a) $(FBC_y,PBC_x)=$ cyclic, (b) $(FBC_y,TPBC_x)=$
M\"obius, (c) $(PBC_y,PBC_x)=$ toroidal, and (d) $(PBC_y,TPBC_x)=$ Klein
bottle.}}
\label{fig1}
\end{figure}

  In the present work we report exact calculations of $P(G,q)$, $W(\{G\},q)$,
and ${\cal B}$ for families of triangular lattice strips of fixed width $L_y=3$
vertices and arbitrary length $L_x=m$ vertices.  The longitudinal and
transverse directions are taken to be $\hat x$ and $\hat y$).  In Fig. 1 we
display some illustrative examples.  
A generic form for chromatic polynomials for recursively defined families of
graphs, of which strip graphs $G_s$ are special cases, is
\beq
P((G_s)_m,q) =  \sum_{j=1}^{N_\lambda} c_j(q)(\lambda_j(q))^m
\label{pgsum}
\eeq
where $c_j(q)$ and the $N_\lambda$ terms $\lambda_j(q)$ depend on the type of
strip graph $G_s$ but are independent of $m$.

   Let us comment on the motivations for the current study and how it extends
previous work.  Calculations of the chromatic polynomials for the cyclic and
M\"obius strip graphs of the square lattice were carried out for $L_y=2$ in
\cite{bds} (see also \cite{bm,b}).  Subsequently, these calculations were
extended to the width $L_y=3$ for the cyclic \cite{wcy} and M\"obius \cite{pm}
square lattice strips. After studies of chromatic zeros in \cite{bkw79,bkw}
and, for $L_y=2$ strips in \cite{bds,readcarib88,read91}, the $W$ and ${\cal
B}$ defined in the infinite-length limit were determined for $L_y=2$ in
\cite{w} and for $L_y=3$ in \cite{wcy,pm}.  Similar calculations were carried
out for strips of various lattices with widths up to $L_y=4$ and free boundary
conditions in \cite{strip,strip2,hs,w2d}, and for cyclic and M\"obius graphs
involving homeomorphic expansions of square strips \cite{pg} and cyclic polygon
chains \cite{nec}.  An important question concerns the effect of boundary
conditions (BC's), and hence graph topology, on $P$, $W$, and ${\cal B}$. We
use the symbols FBC$_y$ and PBC$_y$ for free and periodic transverse boundary
conditions and FBC$_x$, PBC$_x$, and TPBC$_x$ for free, periodic, and twisted
periodic longitudinal boundary conditions.  The term ``twisted'' means that the
longitudinal ends of the strip are identified with reversed orientation.  These
strip graphs can be embedded on surfaces with the following topologies: (i)
(FBC$_y$,FBC$_x$): open strip; (ii) (PBC$_y$,FBC$_x$): cylindrical; (iii)
(FBC$_y$,PBC$_x$): cylindrical, denoted cyclic here; (iv) (FBC$_y$,TPBC$_x$):
M\"obius; (v) (PBC$_y$,PBC$_x$): torus; and (vi) (PBC$_y$,TPBC$_x$): Klein
bottle.\footnote{\footnotesize{These BC's can all be implemented in a manner
that is uniform in the length $L_x$; the case (vii) (TPBC$_y$,TPBC$_x$) with
the topology of the projective plane requires different identifications as
$L_x$ varies and will not be considered here.}}  (We recall that that unlike
graphs of type (i)-(v), the Klein bottle surface cannot be embedded without
self-intersection in ${\mathbb R}^3$.)

Several specific effects of different boundary conditions have been found (for
a general discussion, see \cite{bcc,tw99}).  First, from comparisons of exact
calculations of $W$ for the infinite-length limit of lattice strips with
boundary conditions of types (i)-(iv), it was observed that those with a global
circuit\footnote{\footnotesize{A global circuit is a route following a lattice
direction which has the topology of $S^1$ and a length $\ell_{g.c.}$ that goes
to infinity as $n \to \infty$.}}  lead to a locus ${\cal B}$ that separates the
$q$ plane into different regions, and it was inferred that the presence of
global circuits is a sufficient condition for ${\cal B}$ to have this property
\cite{strip,hs,pg,nec,wcy,pm}.\footnote{\footnotesize{This is not a necessary
condition, as was shown in \cite{strip2}.}}  For strip graphs of regular
lattices, this is equivalent to having $PBC_x$ or $TPBC_x$, i.e., periodic or
twisted periodic boundary conditions in the direction in which the strip length
goes to infinity as $L_x \to \infty$.  Further support for this inference was
adduced from the calculation of $P$, $W$, and ${\cal B}$ for the $L_y=3$ strip
of the square lattice with torus and Klein bottle boundary conditions
\cite{tk}.  Applications of a coloring matrix method \cite{b} for these
calculations has been discussed in \cite{matmeth,bgen}.  For the (homogeneous)
strip graphs with (FBC$_y$,FBC$_x$) in \cite{strip,hs}, ${\cal B}$ consists of
arcs that do not separate regions of the $q$ plane.  As the width of the strip
increases, these arcs tend to elongate, and their ends tend to move toward each
other, thereby suggesting that if one considered the sequence of strip graphs
of this type with width $L_y$ (having taken the limit $L_x \to \infty$ to
obtain a locus ${\cal B}$ for each member of this sequence), then in the limit
$L_y \to \infty$, the arcs would close to form a ${\cal B}$ that separated the
$q$ plane into two or more regions and passed through $q=0$.  The interesting
feature of the families of graphs with global circuits is that when the graphs
contain a global circuit, this separation of the $q$ plane into regions (with
${\cal B}$ passing through $q=0$) already occurs for finite $L_y$. This means
that the $W$ functions of the infinite-length limit of these graphs with cyclic
and twisted cyclic longitudinal boundary conditions already exhibit a feature
which is expected to occur for the ${\cal B}$ for the $W$ function of the full
two-dimensional lattice.  This expectation is supported by the calculation of
$W$ and ${\cal B}$ for the 2D triangular lattice with cylindrical boundary
conditions by Baxter \cite{baxter}.  Thus, although calculations of chromatic
polynomials for lattice strips with global circuits are, in general, more
difficult than for the corresponding strips with free boundary conditions, such
calculations are worthwhile, since in the $L_x \to \infty$ limit, the resultant
locus ${\cal B}$ embodies the analytic property of ${\cal B}$ expected for the
full 2D lattice, viz., that it crosses the real axis at least at $q=0$ and a 
maximal point $q_c$, and separates the $q$ plane into different regions.

Second, it has been found that for a given type of strip graph $G_s$ with
FBC$_y$, the chromatic polynomials for PBC$_x$ and TPBC$_x$ boundary conditions
(i.e., cyclic and M\"obius strips) have the same $\lambda_j$, although in
general different $c_j$.  It follows that the loci ${\cal B}$ are the same for
these two different longitudinal boundary condition choices \cite{pg,wcy,pm}.
In the case of PBC$_y$, the reversal of orientation involved in going from
PBC$_x$ to TPBC$_x$ longitudinal boundary conditions (i.e. from torus to Klein
bottle topology) can lead to the removal of some of the $\lambda_j$'s that were
present; i.e., $P$ for the (PBC$_y$,TPBC$_x$) strip may involve only a subset
of the $\lambda_j$'s that are present for the (PBC$_y$,PBC$_x$) strip.  For
example, for the $L_y=3$ strips of the square lattice with (PBC$_y$, PBC$_x$)
boundary conditions, there are $N_\lambda=8$ $\lambda_j$'s, but for the strip
with (PBC$_y$, TPBC$_x$) boundary conditions only a subset of $N_\lambda=5$ of
these terms occurs in $P$ \cite{tk}.  None of the three $\lambda_j$'s that are
absent from $P$ in the TPBC$_x$ case is leading, so that ${\cal B}$ is the same
for both of these families.

Third, for a given type of strip graph $G_s$ containing a global circuit, it
has been found that, in the infinite-length limit where the locus ${\cal B}$ is
defined, it not only passes through the origin, $q=0$, but always crosses the
positive real axis at one or more points, the maximal one being denoted
$q_c(\{G\})$, as mentioned above.  In contrast, for strip graphs that do not
contain global circuits, ${\cal B}$ may not cross the real axis.  For example,
for the (infinite-length limit of the) strip of the triangular lattice with
free transverse and longitudinal boundary conditions, $(FBC_y,FBC_x)$, the
locus ${\cal B}$ crosses the real axis for width $L_y=3$ but does not cross it
for $L_y=4$ (see Fig. 5 of \cite{strip}).  For square strips with
$(FBC_y,(T)PBC_x)$ boundary conditions, it has been found that $q_c(\{G\})$ is
a nondecreasing function of $L_y$ for the cases studied so far, namely $q_c=2$
for $L_y=1$ and $L_y=2$, and $q_c \simeq 2.33654$ for $L_y=3$
\cite{wcy,pm,bcc}.  For the strips of the triangular lattice with
$(FBC_y,(T)PBC_x)$ boundary conditions, we only have one width for which ${\cal
B}$ has been determined, namely, $L_y=2$, for which $q_c=3$.  It is thus of
interest to explore what the value of $q_c$ is for cyclic and M\"obius strips
of the triangular lattice with larger widths.  For the $L_y=3$ torus and Klein
bottle strips of the square lattice, very interestingly, $q_c=3$, which is the
same value as for the 2D square lattice \cite{tk}.  Clearly it is worthwhile
to investigate what the value of $q_c$ is for torus and Klein bottle strips of
the triangular lattice, and we shall do this here.  

Fourth, for the strip graph $(G_s)_m$ with a given type of transverse boundary
conditions BC$_y$, the chromatic polynomial for PBC$_x$ has a larger value of
$N_\lambda$ than the chromatic polynomial for FBC$_x$, and the corresponding
loci ${\cal B}$ are different.

Fifth, one may ask how the $W$ functions in region $R_1$ compare for 
different boundary conditions.  It has been found \cite{pg,nec,wcy,pm,bcc}
that, for a given type of strip graph $G_s$, in the region $R_1$ (with its
left-hand boundary on the positive real $q$ axis at $q_c$) being 
defined for the PBC$_x$ boundary conditions, the $W$ function is the same for
FBC$_x$, PBC$_x$, and TPBC$_x$.  This includes the region of real $q$ greater
than this value of $q_c$, so that this result is somewhat reminiscent of the
statement of the existence of the thermodynamic limit in statistical mechanics,
i.e. the fact that in the disordered phase of a statistical mechanical 
system the thermodynamic functions are independent of the boundary conditions
and if an ordered phase exists, then the role of the boundary conditions is
only to set a preferred direction for the symmetry-breaking order parameter.

Our new results on $P$, $W$, and ${\cal B}$ for the $L_y=3$ strip of the
triangular lattice with the various boundary conditions of type (iii)-(vi) give
insight concerning these five items, as do our additional results for $L_y=4$
cyclic strips and $L_y=5,6$ cylindrical strips. 
Calculations of these quantities
for the $L_y=2$ cyclic and M\"obius triangular strips were presented in
\cite{wcy}.  An early study of $P$ for the $L_y=2$ strip is in
\cite{sands}\footnote{\footnotesize{We note the following corrections in Sand's
result for $P$ for the $L_y=2$ cyclic strip of the triangular lattice, given on
p. 88 of \cite{sands}: the coefficients of the last two terms should be
reversed in sign to read ($z \equiv q$ in his notation) \ $(z-1)\Bigl [
[(1/2)(5-2z + \sqrt{9-4z})]^m + [(1/2)(5-2z - \sqrt{9-4z})]^m \Bigr ]$, as in
eq. (5.23) of \cite{wcy}.}} and a coloring matrix method to obtain $P$ has been
given in \cite{matmeth}.  $P$, $W$, and ${\cal B}$ have been calculated for
triangular lattice strips with free boundary conditions (i) for $L_y$ up to 4
\cite{strip}. (Some related calculations including work on cases with
noncompact ${\cal B}$, are \cite{read91,w}, \cite{wa}-\cite{sokal}.)

For strips with global circuits, and $L_x$ above the lowest values involving
degenerate cases, the triangular lattice strips with boundary conditions of
type (i)-(vi) have $n=L_yL_x$ vertices; the cyclic and M\"obius triangular
strips have $e=(3L_y-2)L_x$ edges while the torus and Klein bottle triangular
strips have $e=3L_yL_x$ edges.  Define $N_t$ to be the number of triangles in
the strips (i.e., excluding the triangles forming the cross sections of the
strips in the case $L_y=3$); then, again for $L_x$ and $L_y$ large enough to
avoid degenerate cases, the cyclic and M\"obius strips have $N_t=2(L_y-1)L_x$
such triangles, while the torus and Klein bottle strips have $2L_yL_x$ such
triangles.  For strips of type (i), $e = 3L_xL_y-2L_x-2L_y+1$ and
$N_t=2(L_y-1)(L_x-1)$ while for strips of type (ii), $e=3L_xL_y-2L_y$ and
$N_t=2L_y(L_x-1)$.

  Again, for $L_x=m$ and $L_y$ large enough to
avoid degenerate cases, the chromatic number for the triangular
lattice strips with cyclic or torus boundary conditions is
\beq
\chi(tri(L_y)_m,FBC_y,PBC_x) = \chi(tri(L_y)_m,PBC_y,PBC_x) = 
\cases{ 3 & if $m=0$ \ mod 3 \cr
        4 & if $m=1$ or 2 \ mod 3} 
\label{chitricyclictorus}
\eeq
and for the triangular lattice strips with M\"obius or Klein bottle boundary
conditions, 
\beq
\chi(tri(L_y)_m,FBC_y,TPBC_x) = \chi(tri(L_y)_m,PBC_y,TPBC_x) = 4 \quad \forall
\quad m \ge 3 \ . 
\label{chitrimbklein}
\eeq
For strips with $(FBC_y,FBC_x)$ or $(PBC_y,FBC_x)$, $\chi=3$. 

\section{$L_y=3$ Cyclic and M\"obius Triangular Strips} 

We calculate the chromatic polynomials by iterated use of the
deletion-contraction theorem, via a generating function approach 
\cite{strip,hs}.  In general, for a strip graph of type $G_s$ we have 
\beq
\Gamma(G_s,q,x) = \sum_{m=m_0}^{\infty}P((G_s)_m,q)x^{m-m_0}
\label{gamma}
\eeq
where, as before \cite{wcy}, we take $m_0=2$ for strips with periodic or
twisted periodic longitudinal boundary conditions. 
The generating functions $\Gamma(G_s,q,x)$ are rational functions of the form
\beq
\Gamma(G_s,q,x) = \frac{{\cal N}(G_s,q,x)}{{\cal D}(G_s,q,x)}
\label{gammagen}
\eeq
with
\beq
{\cal N}(G_s,q,x) = \sum_{j=0}^{d_{\cal N}} A_{G_s,j}(q) x^j
\label{n}
\eeq
and
\beq
{\cal D}(G_s,q,x) = 1 + \sum_{j=1}^{d_{\cal D}} b_{G_s,j}(q) x^j 
\label{d}
\eeq
where the $A_{G_s,i}$ and $b_{G_s,i}$ are polynomials in $q$ (with no common
factors) and
\beq
d_{\cal N} = deg_x({\cal N})
\label{dn}
\eeq
and
\beq
d_{\cal D} = deg_x({\cal D}) \ . 
\label{dd}
\eeq
The denominator ${\cal D}$ can be written in factorized form as 
\beq
{\cal D}(G_s,q,x) = \prod_{j=1}^{d_{\cal D}}(1-\lambda_{G_s,j}x) \ . 
\label{lambdaform}
\eeq
Equivalently, the $\lambda_{G_s,j}$ are roots of the equation
\beq
\xi^{d_{\cal D}}{\cal D}(G_s,q,1/\xi) = \xi^{d_{\cal D}} +
\sum_{j=1}^{d_{\cal D}} b_{G_s,j}\xi^{d_{\cal D}-j} \ . 
\label{xieq}
\eeq
 From the generating function, we calculate the chromatic polynomials in the
 form of eq. (\ref{pgsum}) by the use of the general formula, eq. (2.14), in 
\cite{hs},
\beq
P(G_m,q) = \sum_{j=1}^{d_{\cal D}} \Biggl [ \sum_{s=0}^{d_{\cal N}}
A_s \lambda_j^{d_{\cal D}-s-1} \Biggr ]
\Biggl [ \prod_{1 \le i \le d_{\cal D}; \ i \ne j}
\frac{1}{(\lambda_j-\lambda_i)} \Biggr ] \lambda_j^{m-m_0}
\label{chrompgsumlam}
\eeq
(where the conventions in eq. (\ref{gamma}) were chosen such that $m_0=0$ in 
\cite{hs}).  Note that some of the coefficients $c_{G_s,j}$ may vanish, so that
not all of the $\lambda_{G_s,j}$'s in ${\cal  D}(G_s,q,x)$ contribute to the
sum in (\ref{pgsum}) \cite{wcy}.  The formula  (\ref{pgsum}) for the chromatic
polynomial has the virtue of being a closed-form expression.  However, for a
given type of strip graph, for values of the width $L_y$ greater than the first
one or two values, the set of $\lambda_j$'s include nonpolynomial algebraic
roots.  These occur, of course, as symmetric functions of the roots,
so that, by a theorem on symmetric functions of roots of algebraic equations,
their resultant contributions to $P$ are expressible in terms of the
coefficients of these equations (which are polynomials in $q$) and hence are 
polynomials in $q$ \cite{pm}. However, in calculating chromatic polynomials
and the resultant chromatic zeros for large values of $L_x=m$, it can be 
convenient to use directly the expression for the generating function, since
both the numerator and denominator of this function, eqs. (\ref{n}), (\ref{d}),
are polynomials in $q$ so that one does not have to rely upon cancellations of
nonpolynomial algebraic expressions at intermediate stages in the calculation.

\vspace{5mm}

The coefficients $c_{tri(L_y),j}$ that enter into the expressions for the
chromatic polynomial (\ref{pgsum}) for the cyclic triangular ($t$) strip of 
width $L_y$ are (like those for the cyclic square strip) certain polynomials 
that we denote $c^{(d)}$, given by \cite{cf}
\beq
c^{(d)} = \prod_{k=1}^d (q-q_{d,k})
\label{cd}
\eeq
where 
\beq
q_{d,k} \equiv 2+2\cos \Bigl ( \frac{2\pi k}{2d+1} \Bigr ) 
\quad {\rm for} \quad
k=1,2,...d
\label{cdzeros}
\eeq
with $0 \le d \le L_y$. We list below the specific $c^{(d)}$'s that 
appear in our results for the $L_y=3$ and $L_y=4$ cyclic strips of the 
triangular lattice 
\beq
c^{(0)}=1 \ , \quad c^{(1)}=q-1 \ , \quad c^{(2)}=q^2-3q+1 \ , 
\label{cd012}
\eeq
\beq
c^{(3)}=q^3-5q^2+6q-1 \ , 
\label{cd3}
\eeq
and
\beq
c^{(4)}=(q-1)(q^3-6q^2+9q-1) \ . 
\label{cd4}
\eeq
Where the coefficient $c^{(0)}$ appears in chromatic polynomials, we shall
simply write it as unity. 

\vspace{8mm}

For the cyclic $L_y=3$ triangular strip of length $L_x=m$, we obtain the 
result 
\beqs
& & P(tri(3 \times m),FBC_y,PBC_x,q)=\sum_{j=1}^{10} c_{t3,j} 
(\lambda_{t3,j})^m
\cr\cr
& = & c^{(3)}(-1)^m 
+ c^{(2)}\Bigl [ (q-2)^m+(\lambda_{t3,3})^m + (\lambda_{t3,4})^m \Bigr ] 
\cr\cr
& + & c^{(1)}\Bigl [ [-(q-2)(q-3)]^m + (\lambda_{t3,6})^m + (\lambda_{t3,7})^m
 + (\lambda_{t3,8})^m \Bigr ] + \Bigl [ 
(\lambda_{t3,9})^m + (\lambda_{t3,10})^m  \Bigr ] \ . \cr\cr 
& & 
\label{ptcyclic}
\eeqs
where the short notation $t3$ is used in subscripts for $tri(L_y=3)$.  
The labelling of $c_{t3,j}$ and $\lambda_{t3,j}$ in eq. (\ref{ptcyclic}) is 
consecutive, so that $c_{t3,1}=c^{(3)}$, $\lambda_{t3,1}=-1$; 
$c_{t3,2}=c^{(2)}$, $\lambda_{t3,2}=q-2$, and so forth: 
\beq
\lambda_{t3,(3,4)}=\frac{1}{2}\Bigl ( -7+2q \pm \sqrt{25-8q} \ \Bigr ) \ , 
\label{lamt34}
\eeq
$\lambda_{t3,j}$, $j=6,7,8$ are the roots of the cubic equation 
\beq
\xi^3+b_{t3,1}\xi^2+b_{t3,2}\xi + b_{t3,3}=0 
\label{cubeq}
\eeq
where 
\beq
b_{t3,1}=2q^2-12q+19
\label{bt31}
\eeq
\beq
b_{t3,2}=(q-2)(q^3-9q^2+28q-29)
\label{bt32}
\eeq
\beq
b_{t3,3}=-(q-2)^4(q-3) \ , 
\label{bt33}
\eeq
and 
\beq
\lambda_{t3,(9,10)}=\frac{1}{2}\biggl [ q^3-7q^2+18q-17 \pm 
(q^6-14q^5+81q^4-250q^3+442q^2-436q+193)^{1/2} \biggr ] \ . 
\label{lamt910}
\eeq
Note that some $\lambda_{t3,j}$'s vanish if $q=2$ or $q=3$; specifically, if 
$q=2$, then $\lambda_{t3,3}=0$ and two of the roots of the cubic equation
(\ref{cubeq}) vanish, while for $q=3$, $\lambda_{t3,3}=0$ and one of the 
roots of eq. (\ref{cubeq}) vanish.  The vanishing of the chromatic polynomial
at $q=0,1,2$ and (i) $q=3$ for the cyclic strip with $m \ne 0$ mod 3 or (ii)
$q=3$ for the M\"obius strip discussed below involves both some terms
$\lambda_j$ vanishing and cancellations among others.  This is also true of
chromatic polynomials for strips of other lattices. 

Parenthetically, we note that the denominator of the generating function for 
this strip graph actually has $d_{\cal D}=16$:
\beq
{\cal D}(tri(L_y=3),FBC_y,PBC_x,q,x) = \sum_{j=1}^{16}(1-\lambda_{t3,j}(q)x)
\label{lambdaformt}
\eeq
where the $\lambda_{t,j}$ were given in eq. (\ref{ptcyclic}) and 
\beq
\lambda_{t,11}=\lambda_{t,1}=-1
\label{lamt11}
\eeq
\beq
\lambda_{t,12}=\lambda_{t,13}=q-3
\label{lamt1213}
\eeq
and $\lambda_{t,j}$, $j=14,15,16$ are the roots of the cubic equation 
\beq
\xi^3+(2q-5)(q-3)\xi^2+(q-2)^2(q-3)(q-4)\xi-(q-2)^4(q-3)=0 \ . 
\label{extracubic}
\eeq 
A new feature of these results that we have not encountered before in our
calculations of chromatic polynomials and generating functions for strip graphs
is the occurrence of factors in ${\cal D}$ that have multiplicity higher than
unity: specifically, the factors $(1+x)^2$, corresponding to
$\lambda_{t3,1}=\lambda_{t3,11}=-1$, and $[1-(q-3)x]^2$, corresponding to
$\lambda_{t3,12}=\lambda_{t3,13}=q-3$.  None of the $\lambda_{t3,j}$, 
$12 \le j \le 16$, contribute to $P$ (and $\lambda_{t3,11}$ is identical to 
$\lambda_{t3,1}$), so that $P$ only involves the ten $\lambda_{t3,j}$'s in
eq. (\ref{ptcyclic}).  We had found cases earlier of some $\lambda_j$'s in
${\cal D}$ not contributing to $P$, e.g., for the $L_y=2$ cyclic strip of the
kagom\'e lattice and homeomorphic expansions of the $L_y=2$ cyclic strip of the
square lattice \cite{wcy}.  However, this is the first value of $L_y$ where we
have encountered this type of behavior for a strip of a homopolygonal
lattice.\footnote{\footnotesize{As in \cite{cmo,wn}, we define a homopolygonal
2D lattice as a regular tiling of the plane involving a single type of polygon,
in which all vertices are equivalent, while a heteropolygonal lattice is the
same but involving two or more different types of regular polygons. In the case
of the kagom\'e lattice, these are triangles and hexagons.}}

The numerator of the generating function has degree 
$deg_x({\cal N})=14$.  The polynomials $A_j$, $j=1,.., 14$ comprising this
numerator can be calculated from the expressions for $P$ and ${\cal D}$, 
eqs. (\ref{ptcyclic}) and (\ref{lambdaformt}) and hence are not listed here. 

 From eq. (\ref{chitricyclictorus}), it follows that for $m=1$ or 2 mod 3 and 
$q=3$, the chromatic polynomial in eq. (\ref{ptcyclic}) vanishes; for $m=0$ 
mod 3 (and $m \ge 3$) and $q=3$ we find 
\beq
P(tri(L_y=3)_m,FBC_y,PBC_x,q=3)=6 \quad {\rm for} \quad m = 0 \quad mod \ 3 \ .
\label{ptq3m0mod3}
\eeq
This may be contrasted with the case of a $k$-critical graph $G_k$, where 
$\chi(G)=k$ and $P(G,q=\chi(G))=k!$ (e.g. $k=3$ for the triangular lattice). 

We define
\beq
C(G)=\sum_{j=1}^{N_{\lambda_G}} c_{G,j} \ . 
\label{cgsum}
\eeq
where the $G$-dependence in the coefficients is indicated explicitly.  Note
that for recursive graphs like the strip graphs considered here, the $c_{G,j}$
depend on $L_y$ and the boundary conditions, but not on $L_x$. Our results 
above give 
\beq
C=q(q-1)^2 \quad {\rm for} \quad G=tri(L_y=3,FBC_y,PBC_x) \ .
\label{csumtrily3cyc}
\eeq
in accord with the generalization \cite{wcy,pm} $C=P(T_{L_y},q)$ for this
strip, where $P(T_n,q)=q(q-1)^{n-1}$ and $T_n$ is the tree graph with $n$
vertices.  This was also true for the $L_y=3$ cyclic strip of the square
lattice \cite{wcy,pm} and is in accord with the coloring matrix approach
\cite{tk,matmeth}.

\vspace{6mm}

For the $L_y=3$ triangular lattice M\"obius strip (denoted by the subscript 
$t3Mb$) we calculate 
\beq 
P(tri(L_y=3)_m,FBC_y,TPBC_x,q) = \sum_{j=1}^{10}c_{t3Mb,j} (\lambda_{t3Mb,j})^m
\label{pmobius}
\eeq
where 
\beq
\lambda_{t3Mb,j} = \lambda_{t3,j} \ , \quad j=1,..,10 \ . 
\label{lamtm}
\eeq
As was true in the case of the L$_y=2$ M\"obius strip of the triangular 
lattice, the coefficients $c_j$ are not just of the $c^{(d)}$ polynomial 
form; instead, some coefficients are rational functions of $q$ and others are 
algebraic nonpolynomial functions of $q$: 
\beq
c_{t3Mb,1}=c^{(2)}
\label{c1tmb}
\eeq
\beq
c_{t3Mb,2}=-\frac{(q^2-4q+5)}{2(q-2)}
\label{c2tmb}
\eeq

\beq
c_{t3Mb,(3,4)} = \frac{1}{4(q-2)} \biggl [ (q-3)^2 \mp \frac{(q-1)(q-5)}
{\lambda_{t,3}-\lambda_{t,4}} \ \biggr ]
\label{c34tmb}
\eeq

\beq
c_{t3Mb,5} = \frac{(q-1)(q^2-4q+5)}{2(q-2)}
\label{c5tmb}
\eeq

\beq
c_{t3Mb,9}=c_{t3Mb,10}=c^{(0)}=1 \ . 
\label{c910tmb}
\eeq
The $j=3,4$ terms can be written in a manner which manifestly 
exhibits their property of being a symmetric function of $\lambda_{t3,3}$ and 
$\lambda_{t3,4}$:
\beq
c_{t3Mb,3}(\lambda_{t3,3})^m+c_{t3Mb,4}(\lambda_{t3,4})^m = 
\frac{(q-3)^2}{4(q-2)}\Bigl [ (\lambda_{t3,3})^m+(\lambda_{t3,4})^m \Bigr ] 
-\frac{(q-1)(q-5)}{4(q-2)}\biggl [ \frac{(\lambda_{t3,3})^m-(\lambda_{t3,4})^m}{
\lambda_{t3,3}-\lambda_{t3,4}} \biggr ] \ . 
\label{ctm34}
\eeq
It is convenient to leave the expressions for $c_{t3Mb,j}$, $j=6,7,8$, in the
general forms that follow from eq. (\ref{chrompgsumlam}):
\beq
c_{t3Mb,j}=\lambda_{t3,j}^{-2} \Biggl [ \sum_{s=0}^8 A_{tm3,s} 
(\lambda_{t3,j})^{9-s} \Biggr ]\Biggl [ \prod_{1 \le i \le 10; \ i \ne j}
\frac{1}{(\lambda_{t3,j}-\lambda_{t3,i})} \Biggr ] \ , \quad j=6,7,8
\label{ctmj}
\eeq
where the $A_{t3Mb,s}$ are given in the appendix. 

For the sum of the coefficients for the $L_y=3$ M\"obius strip of the 
triangular lattice we calculate 
\beq
C=\sum_{j=1}^{10}c_{t3Mb,j}=q(q-1) \quad {\rm for} \quad
G=tri(L_y=3,FBC_y,TPBC_x)
\label{csumtrily3mb}
\eeq
in agreement with a general formula that we have derived elsewhere for a
M\"obius strip graph $G_s$ of width $L_y$ of the square or triangular lattice 
\cite{cf},
\beq
C(G_s(L_y),Mb)=\cases{ 0 & for even $L_y$ \cr
P(T_{(\frac{L_y+1}{2})},q) & for odd $L_y$ \cr } \ . 
\label{csummb}
\eeq

\begin{figure}[p]
\centering
\leavevmode
\epsfxsize=4.0in
\begin{center}
\leavevmode
\epsffile{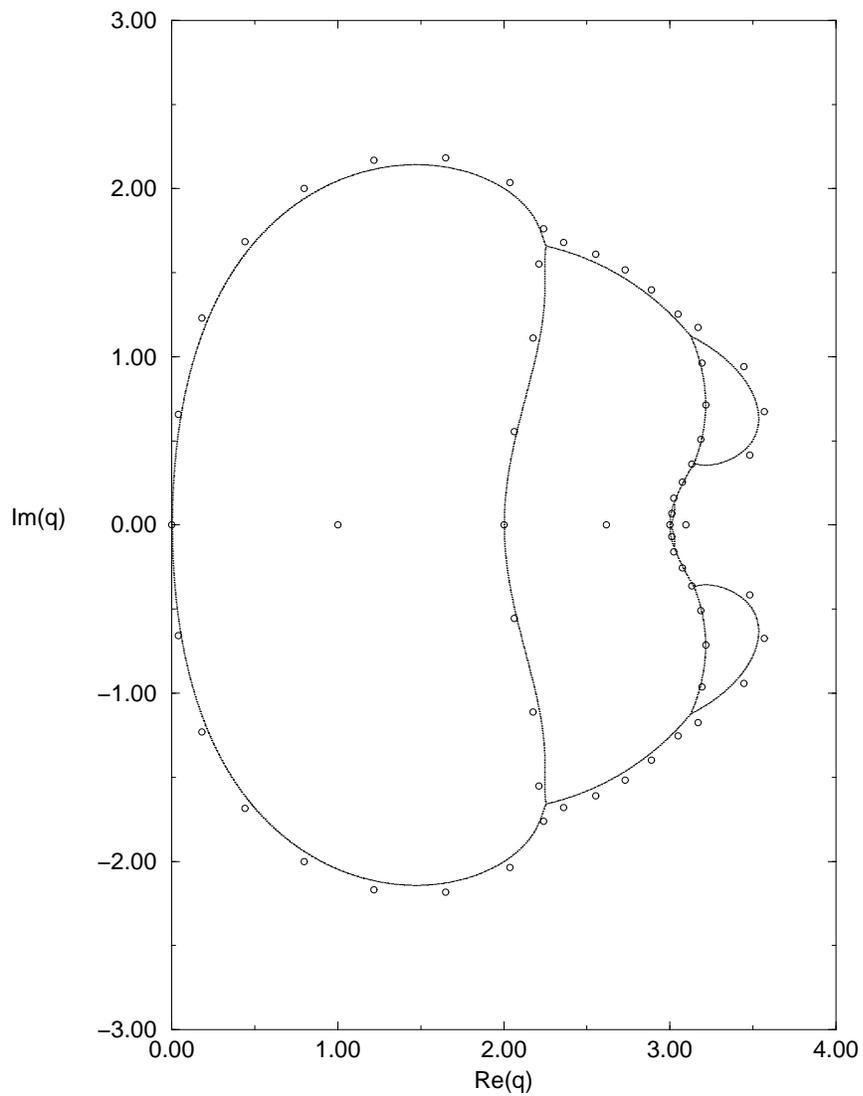}
\end{center}
\caption{\footnotesize{${\cal B}$ and chromatic zeros for the $L_y=3$,
$L_x=m=20$ (i.e., $n=60$) cyclic triangular strip graph.}}
\label{tricyczeros}
\end{figure}

Chromatic zeros for the $L_y=3$, $L_x=m=20$ cyclic graph of the triangular
lattice are shown in Fig.  \ref{tricyczeros}; with this value of $m$, the
chromatic zeros for the triangular lattice M\"obius graph are quite similar and
both sets of chromatic zeros lie close to the boundary ${\cal B}$ and indicate
its position.  The locus ${\cal B}$ and the $W$ functions are the same for the
cyclic and M\"obius graph families.  We find
\beq
q_c(tri(L_y=3),FBC_y,(T)PBC_x)=3 \ . 
\label{qctcycly3}
\eeq 
This will be discussed further below when we give our result for $q_c$ for the
corresponding $L_y=4$ strip of the triangular lattice.

The locus ${\cal B}$ has support for $Re(q) \ge 0$ and separates the $q$ plane
into five main regions.  The outermost one, region $R_1$, extends to infinite
$|q|$ and includes the intervals $q \ge 3$ and $q \le 0$ on the real $q$ axis.
The other regions are labelled consequtively moving leftward along the real
axis from $q=q_c=3$ to $q=0$, and then including complex conjugate pairs of
regions that have no overlap with the real axis.  Thus, region $R_2$ includes
the real interval $2 \le q \le 3$, while region $R_3$ includes the real
interval $0 \le q \le 2$.  There are also two complex-conjugate regions, $R_4,
R_4^*$ centered at approximately $q \simeq 3.35 \pm 0.7i$.  The triple points
occur at (i) $q \simeq 3.14+0.37i$ and (ii) $q \simeq 3.12+1.12i$, where $R_1,
\ R_2$, and $R_4$ are contiguous, and (iii) $q \simeq 2.25+ 1.66i$, where
$R_1$, $R_2$, and $R_3$ are contiguous, together with the complex conjugates of
these points.  In addition, there are two tiny regions $R_5, \ R_5^*$ that
touch the real axis at the single point $q_c=3$ and extend a very short
distance to the upper and lower right, with triple points at about $q \simeq
3.02 \pm 0.13i$.  These are plotted in Fig. \ref{tricyczeros}. Thus, $q_c$ is a
multiple point where four branches of ${\cal B}$ intersect.  The occurrence of
extremely small regions was also found for the ($L_x \to \infty$ limit of the)
$L_y=3$ cyclic strip of the square lattice, as shown in Fig. 1 of \cite{wcy}.
The fact that such tiny regions can occur means that when one performs the
usual computer scan over the complex plane to map out the region diagram and
determine the dominant $\lambda_j$'s in each region, since this involves a
finite grid, one can only detect tiny sliver regions down to a certain
resolution set by the grid of the scan.  On the real axis, this is not a
serious complication, since the mapping is a one-variable problem, but in the
complex plane, one must a commensurately fine scanning grid to detect tiny
sliver regions.

The part of ${\cal B}$ and the associated chromatic zeros with largest
magnitude occur at roughly $q \simeq 3.53 + 0.69i$ and have $|q| \simeq 3.59$.
Comparing the present locus ${\cal B}$ for $L_y=3$ with the one obtained in
\cite{wcy} for $L_y=2$, we obtain further evidence supporting the conjecture
that for a strip graph of a regular lattice with periodic or twisted periodic
longitudinal boundary conditions (${\cal B}$ is the same for both of these),
the envelope curve of ${\cal B}$, i.e. the outermost portion of ${\cal B}$ for
a given $L_y$ surrounds the envelope curve for the ${\cal B}$ of the same strip
with a smaller value of $L_y$.  This was found to be true for the cyclic and
M\"obius strips of the square lattice for the cases $L_y=2,3$ and 4 (and, for
the cyclic case, also $L_y=1$) \cite{w,wcy,bcc}.  For strips of a regular
lattice $\Lambda$, as $L_y \to \infty$, one expects that the envelope curves
defined for each $L_y$ will approach a limiting curve, which is precisely the
envelope for the boundary ${\cal B}$ of the full 2D lattice $\Lambda$ defined
via this limit.

In region $R_1$, $\lambda_{t3,9}$ is the dominant $\lambda_j$ (with appropriate
choice of branch cut for evaluation away from the negative real axis), so 
\beq
W=(\lambda_{t3,9})^{1/3} \ , \quad q \in R_1 \ . 
\label{wt3r1}
\eeq
The fact that this is the same as $W$ for the (FBC$_y$,FBC$_x$) case 
\cite{strip} is a general result \cite{bcc}.  
In region $R_2$, $\lambda_{t3,4}$ is dominant, so
\beq
|W| = |\lambda_{t3,4}|^{1/3} \ , \quad q \in R_2
\label{wt3r2}
\eeq
(in regions other than $R_1$, only $|W|$ can be determined unambiguously
\cite{w}). In region $R_3$, $W$ is given by the largest (in magnitude) of the 
roots of the cubic (\ref{cubeq}), which we label $\lambda_{t3,6}$:
\beq
|W|=|\lambda_{t3,6}|^{1/3} \ , \quad q \in R_3 \ .
\label{wt3r3}
\eeq
In regions $R_4, R_4^*$, 
\beq
|W|=|\lambda_{t3,8}|^{1/3} \ , \quad q \in R_4 \ , R_4^* \ .
\label{wt3r4}
\eeq
where $\lambda_{t3,8}$ is the other among these cubic roots that has maximal
magnitude in this region.  In $R_5, \ R_5^*$, $|W|=|\lambda_{t3,6}|^{1/3}$.

\section{$L_y=4$ Cyclic Triangular Strip}

We have succeeded in performing an exact calculation of the chromatic
polynomial for the cyclic triangular strip of arbitrarily great length and of
the next larger width, namely, $L_y=4$.  
The calculation is considerably more involved than for the $L_y=3$
cyclic strip, as is indicated by the number of $\lambda_j$ terms in
eq. (\ref{pgsum}), namely, $N_\lambda=26$, as contrasted with the value
$L_y=10$ for the $L_y=3$ cyclic strip.  Elsewhere we have proved that 
$N_\lambda$ is the same for the square and triangular strips of a given width 
and have given a general determination of $N_\lambda$ as a function of $L_y$ 
\cite{cf}.  To go beyond this width for this family of cyclic strips of the 
triangular lattice would be increasingly difficult, since the number of
terms $N_\lambda$ in (\ref{pgsum}) is 70, \ 192, \ and 534 for $L_y=5, \ 6,$ 
and 7, respectively. For $L_y=4$ we find the chromatic polynomial
\beq 
P(tri(4 \times m), FBC_y,PBC_x,q)=\sum_{j=1}^{26}
c_{t4,j}(\lambda_{t4,j})^m
\label{pt4}
\eeq
where one term is $\lambda_{t4,1}=1$ and the others, $\lambda_{t4,j}$, are 
the roots of
(i) the quartic equation (\ref{t4eq1}), for $2 \le j \le 5$; (ii) the quartic
equation (\ref{t4eq2}), for $6 \le j \le 9$; (iii) the degree-8 equation
(\ref{t4eq3}), for $10 \le j \le 17$; and (iv) the degree-9 equation
(\ref{t4eq4}), for $18 \le j \le 26$.  Since these equations are somewhat
lengthy, they are given in the appendix.  In contrast to the $L_y=2$ and
$L_y=3$ cyclic triangular strips, here it is not possible to solve for all of
the terms $\lambda_{t4,j}$ as algebraic roots since for $j=10$ through $j=26$
these are roots of equations of degree higher than quartic.  
The corresponding coefficients 
$c_{t4,j}$ are 
\beq
c_{t4,1}=c^{(4)}
\label{ct4j1}
\eeq
\beq
c_{t4,j}=c^{(0)} \quad {\rm for} \quad 2 \le j \le 5
\label{ct4j25}
\eeq
\beq
c_{t4,j}=c^{(3)} \quad {\rm for} \quad 6 \le j \le 9
\label{ct4j69}
\eeq
\beq
c_{t4,j}=c^{(2)} \quad {\rm for} \quad 10 \le j \le 17
\label{ct4j1017}
\eeq
\beq
c_{t4,j}=c^{(1)} \quad {\rm for} \quad 18 \le j \le 26  \ .
\label{ct4j1826}
\eeq
As before, one can, equivalently, list the generating function; however, since
all of the relevant information is present in (\ref{pt4}) with the requisite
definitions of the coefficients $c_{t4,j}$ and terms $\lambda_{t4,j}$, we shall
forego this. 

For the M\"obius $L_y=4$ strip of the triangular lattice, a general argument
\cite{pm} shows that the $\lambda_j$'s are the same as for the cyclic strip,
although the coefficients $c_j$ are different and are more complicated than
those for the cyclic strip, as was already shown by the $L_y=2$ triangular
strip, where two coefficients for the M\"obius case are algebraic nonpolynomial
functions of $q$.  As discussed before, since ${\cal B}$ is determined only by
the $\lambda_j$'s, it is the same for the cyclic and M\"obius strips of a given
lattice.

\begin{figure}[p]
\centering
\leavevmode
\epsfxsize=4.0in
\begin{center}
\leavevmode
\epsffile{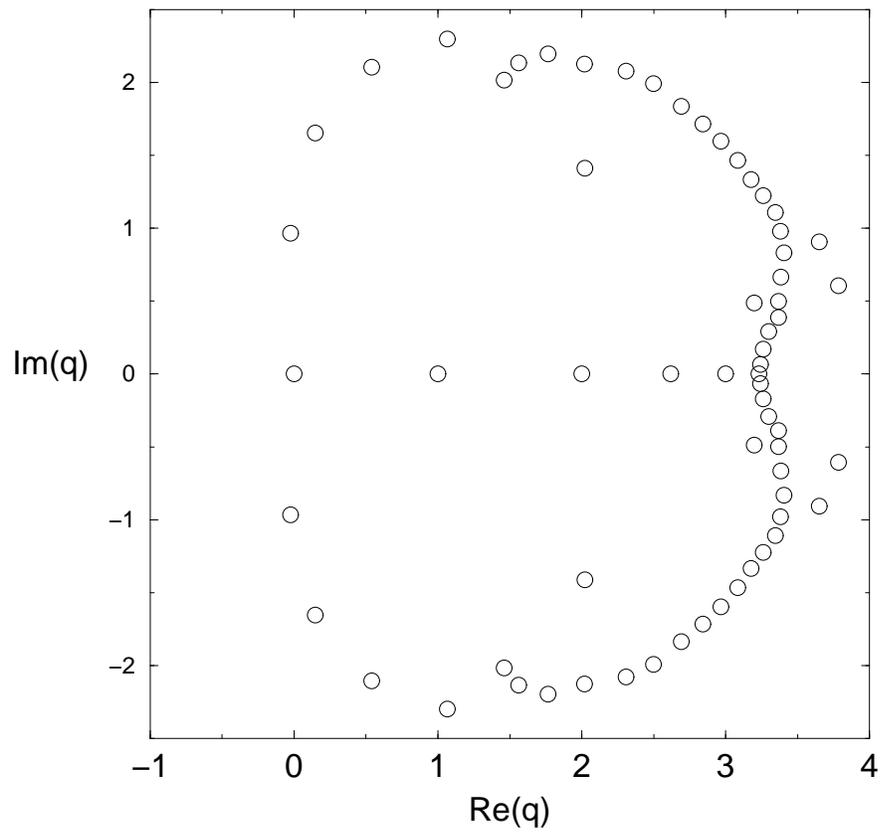}
\end{center}
\caption{\footnotesize{${\cal B}$ and chromatic zeros for the $L_y=4$,
$L_x=m=16$ (i.e., $n=64$) cyclic triangular strip graph.}}
\label{tricyczeros4}
\end{figure}

Chromatic zeros for the $L_y=4$, $L_x=m=16$ (hence $n=64$) cyclic graph of the
triangular lattice are shown in Fig. \ref{tricyczeros4}; the complex zeros give
an approximate indication of ${\cal B}$.  The locus ${\cal B}$ separates the
$q$ plane into six main regions, including four that contain intervals of the
real axis.  The outermost one, region $R_1$, extends to infinite $|q|$ and
includes the intervals $q \ge q_c$ and $q \le 0$ on the real $q$ axis.  Region
$R_2$ includes the real interval $3 \le q \le q_c$; region $R_3$ includes the
real interval $2 \le q \le 3$, and region $R_4$ includes the real interval $0
\le q \le 2$.  There are also complex-conjugate regions $R_5$ and $R_5^*$
centered at approximately $q \simeq 3.5 \pm 0.6i$.  There could also be other
such complex-conjugate pairs of regions.  The density of zeros is observed to
be smaller on the parts of ${\cal B}$ extending through $q=2$ and $q=3$ and
also on the right-most bubble-like curves than on the rest of the of ${\cal
B}$.  As before with the $L_y=3$ strip, it is straightforward to determine the
various triple points.  The maximal point at which the locus ${\cal B}$ crosses
the real axis is \beq q_c(tri(L_y=4),FBC_y,(T)PBC_x) \simeq 3.228126 \ .
\label{qctcycly4}
\eeq 
Thus, for the triangular strips with $(FBC_y,(T)PBC_x)$ boundary
conditions for which the chromatic polynomials have been calculated so far,
i.e. for the widths $L_y=2,3$, and 4, $q_c$ is a nondecreasing function of
$L_y$: $q_c(tri(2))=3$, $q_c(tri(3))=3$, and $q_c(tri(4))=3.228...$.  We have
found that this is also true for the strips of the square lattice with the same
cyclic or M\"obius boundary conditions: $q(sq(1))=q_c(sq(2))=2$, $q_c(sq(3))
\simeq 2.34$, and $q_c(sq(4)) \simeq 2.49$.  This contrasts with the
non-monotonic behavior of $q_c$ as a function of $L_y$ that we have found for
strips with $(PBC_y,FBC_x)$ boundary conditions (see eq. (\ref{qctpy5}) below).
It is anticipated that for cyclic and M\"obius strips of the triangular
lattice, as $L_y$ increases beyond 3, $q_c$ will approach the 2D value
$q_c(tri)=4$.

Another important feature of the locus ${\cal B}$ for the $L_y=4$ cyclic 
triangular strip, which is the same as was true of the $L_y=2$ and $L_y=3$ 
cyclic triangular strips, is that it crosses the real axis at the points 
$q=0,2$ and 3.  This leads us to the conjecture
\beq
Conjecture: \quad {\cal B} \supset \{q=0, \ 2, \ 3 \}  \ \  {\rm for} \ \ 
tri(L_y,FBC_y,(T)PBC_x) \quad \forall \ L_y \ge 2
\label{bcrossq023}
\eeq
The crossing at $q=0$ is present in a very wide class of families of graphs 
that contain global circuits \cite{bcc}. 

The crossing at $q=2$ that we have found for $L_y=2,3,4$ signals a
zero-temperature critical point of the Ising antiferromagnet on these
triangular strips, and similarly for the generalization (\ref{bcrossq023}) to
arbitrary $L_y$.  Note that this $T=0$ critical point involves frustration,
since not all of the spin-spin interactions around a triangle can have their
energies minimized.  Frustration is also responsible for the $T=0$ critical
point of the Ising antiferromagnet on the full 2D triangular lattice
\cite{wannier,stephenson}, but the nature of the critical singularities in the
free energy, correlation length, etc. are different: they are algebraic for the
triangular lattice, but are essential singularities for the $L_y \times \infty$
strips \cite{ta}.  As will be shown below, this crossing of ${\cal B}$ at $q=2$
is also found for the $L_y=2$ strips with torus or Klein bottle boundary
conditions, i.e.  $(PBC_y,PBC_x)$ or $(PBC_y,TPBC_x)$.  Thus, it appears to be
a general feature of the strip graphs of the triangular lattice with periodic
or twisted periodic longitudinal boundary conditions.  Indeed, the respective
boundaries ${\cal B}$ for the $L_y=2$ cyclic and M\"obius strips of the square,
triangular, and kagom\'e lattices, and for the $L_y=3$ cyclic and M\"obius
strips of the square lattice \cite{wcy} and the torus and Klein bottle strips
of the square lattice \cite{tk} also cross the real $q$ axis at $q=2$ (as well
as $q=0$).

In contrast, for strips with $(FBC_y,FBC_x)$ (open) or $(PBC_y,FBC_x)$
(cylindrical) boundary conditions, ${\cal B}$ does not pass through $q=0$ and,
while ${\cal B}$ for the $L_y=3$ open square strip crosses the real axis at
$q=2$, the respective loci ${\cal B}$ for the $L_y=4,5$ open and cylindrical
square strips and the open and cylindrical triangular strips with $L_y=3,4,5$
do not, as discussed further below.

In regions $R_j$, $1 \le j \le 4$, the dominant terms are, respectively, the
root of maximal magnitude of (1) the first quartic equation (\ref{t4eq1}), 
the second quartic equation (\ref{t4eq2}), (3) the eighth-degree equation 
(\ref{t4eq3}), (4) the ninth-degree equation (\ref{t4eq4}).  In region $R_1$, 
we label the dominant term as $\lambda_{t4,R1}$, and so forth for the other
regions. In the complex-conjugate regions $R_5, R_5^*$ 
the dominant term is the largest-magnitude root of the ninth-degree equation 
(\ref{t4eq4}), which we label as $\lambda_{t4,R5}$.  We have 
\beq 
W=(\lambda_{t4,R1})^{1/4} \ , \quad q \in R_1
\label{wt4r1}
\eeq
\beq
|W| = |\lambda_{t4,Rj}|^{1/4} \ , \quad q \in R_j \ , j=2,3,4
\label{wt4r23}
\eeq
and 
\beq
|W| = |\lambda_{t4,R5}|^{1/4} \ , \quad q \in R_5, R_5^* \ . 
\label{wt4r45}
\eeq
As usual, the boundaries of the regions are defined by the degeneracies, in 
magnitude, of the leading $\lambda_{t4,j}$'s in these regions.  For example, 
on the real axis, one has $|\lambda_{t4,R1}|=|\lambda_{t4,R2}|$ at $q_c$, 
$|\lambda_{t4,R2}|=|\lambda_{t4,R3}|$ at $q=3$, $|\lambda_{t4,R3}|=
|\lambda_{t4,R4}|$ at $q=2$, and $|\lambda_{t4,R4}|=|\lambda_{t4,R1}|$ at
$q=0$.  

In contrast to the situation for the $L_y=2$ and $L_y=3$ cyclic strips of the
triangular lattice, here $q_c$ is a regular instead of a multiple (=
intersection) point on the algebraic curve forming ${\cal B}$, in the
terminology of algebraic geometry (a multiple point on an algebraic curve is a
point where several branches of the curve cross each other). 

Another qualitative difference is that for $L_y=2$ and $L_y=3$, ${\cal B}$
contained support only for $Re(q) \ge 0$, and the only point on ${\cal B}$ with
$Re(q)=0$ was the origin itself; however, for $L_y=4$, ${\cal B}$ extends
slightly into the half-plane with $Re(q) < 0$.  Although these lattice strips
involve global circuits, it was found in \cite{hs} that this is not a necessary
condition for chromatic zeros and loci ${\cal B}$ of lattice strips (or
homeomorphic expansions thereof) to have support for $Re(q) < 0$; this was also
evident in results of \cite{baxter}, ruling out a conjecture in \cite{strip}.
In the case of strips of the square lattice we had found that for $L_y=1$ and
2, ${\cal B}$ only had support for $Re(q) \ge 0$, but for $L_y=3$ and $L_y=4$
it had support also for $Re(q) < 0$.

Finally, another general feature is that the motion of the outermost curves
forming part of ${\cal B}$ for $L_y=2,3,4$ is consistent with the inference
that as $L_y$ increases, these form a limiting curve.  

\section{$L_y=3$ Triangular Lattice Strips with Torus and Klein Bottle 
Boundary Conditions}

For the $L_y=3$ triangular lattice strips with torus boundary conditions 
(denoted with the subscript $tt$), we calculate, by the same methods as above, 
\beq
P(tri(L_y=3)_m,PBC_y,PBC_x,q) = \sum_{j=1}^{11} c_{tt,j} (\lambda_{tt,j})^m
\label{ptorus}
\eeq
where
\beq
\lambda_{tt,1}=-2 \ , \qquad c_{tt,1} = \frac{1}{3}(q-1)(q^2-5q+3) \ ,
\label{lamtt1}
\eeq

\beq
\lambda_{tt,2}=q-2 \ , \qquad c_{tt,2}=\frac{1}{2}(q-1)(q-2) \ ,
\label{lamtt2}
\eeq

\beq
\lambda_{tt,3}=3q-14 \ , \qquad c_{tt,3}=\frac{1}{2}q(q-3) \ ,
\label{lamtt3}
\eeq

\beq
\lambda_{tt,4}=-2(q-4)^2 \ , \qquad c_{tt,4}=q-1 \ , 
\label{lamtt4}
\eeq

\beq
\lambda_{tt,5}=q^3-9q^2+29q-32 \ , \qquad c_{tt,5}=1  \ ,
\label{lamtt5}
\eeq

\beq
\lambda_{tt,(6,7)}=e^{\pm 2\pi i/3} \ , \qquad c_{tt,6} = c_{tt,7} = 
\frac{1}{6}q(q-1)(2q-7) \ ,
\label{lamtt67}
\eeq

\beq
\lambda_{tt,(8,9)}=(q^2-5q+7)e^{\pm 2\pi i/3} \ , \qquad 
c_{tt,8}=c_{tt,9}=q-1 \ , 
\label{lamtt89}
\eeq

\beq
\lambda_{tt,(10,11))}=-(2q-7)e^{\pm 2\pi i/3} \ , \qquad c_{tt,10}=c_{tt,11}=
\frac{1}{2}(q-1)(q-2) \ . 
\label{lamtt1011}
\eeq
Note that although several of the coefficients are complex (and have
coefficients that are not integers), the chromatic polynomial itself is, of
course, a polynomial with integer coefficients. 
The sum of the complex terms can be written as
\beq
\sum_{j=6}^{11} c_{tt,j} (\lambda_{tt,j})^m  = 
(q-1)\Bigl [ \frac{q(2q-7)}{3} + 2(q^2-5q+7)^m + (q-2)[-(2q-7)]^m \Bigr ] 
\cos \Bigl (\frac{2m \pi}{3} \Bigr ) 
\label{osctermsum}
\eeq
None of these terms with $6 \le j \le 11$ can be dominant anywhere in the $q$
plane since this would imply that the limit (\ref{w}) would not exist. 
As compared with previously calculated strip graphs, this exhibits a number of
qualitatively new features: (i) previously, only one $\lambda_j$ was a
constant, independent of $q$, and this was always equal to either 1 or $-1$;
here there are three, and none of these is equal to $\pm 1$; (ii) previously,
all of the $\lambda_j$'s were either polynomials with real coefficients or
algebraic functions of polynomials with real coefficients.  By general 
arguments given before \cite{bcc}, the dominant $\lambda_j$ in region $R_1$ has
as its highest power $q^{n/m}$, i.e., in the present case, $q^3$, and has
coefficient 1.  We find that the generating function for this case has 
$deg_x({\cal N})=9$ and $deg_x({\cal D})=11$, so that all of the $\lambda_j$'s
in ${\cal D}$ contribute to $P$. 

Given that $\chi=4$ for $m=1$ or 2 mod 3 from eq. (\ref{chitrimbklein}), it 
follows that for these $m$ values,
and $q=3$, $P(tri(L_y=3)_m,PBC_y,PBC_x,q=3)=0$; however, $\chi=3$ for $m=0$ mod
3; in this case, we find (for $m \ge 3$)
\beq
P(tri(L_y=3)_m,PBC_y,PBC_x,q=3)=6 \quad {\rm for} \quad m = 0 \quad mod \ 3 \ .
\label{pttq3m0mod3}
\eeq
This is analogous to eq. (\ref{ptq3m0mod3}) above. 

\vspace{6mm}

For the $L_y=3$ triangular lattice strips with Klein bottle topology
(denoted with the subscript $tk$) we calculate 
\beqs
& & P(tri(L_y=3)_m,PBC_y,TPBC_x,q) = \sum_{j=1}^5 c_{tk,j} (\lambda_{tk,j})^m 
\cr\cr
& & = -(q-1)(-2)^m-\frac{1}{2}(q-1)(q-2)(q-2)^m 
+ \frac{1}{2}q(q-3)(3q-14)^m + \cr\cr
& & (q-1)[-2(q-4)^2]^m + (q^3-9q^2+29q-32)^m \ . 
\label{ptk}
\eeqs
Thus, $c_{tk,j}=c_{tt,j}$, $j=1,..5$, and the six $\lambda_{tt,j}$, 
$j=6,..11$ that involve complex factors do not contribute to 
$P(tri(L_y=3)_m,PBC_y,TPBC_x,q)$.  Since none of these six $\lambda_{tt,j}$'s
is dominant anywhere, it follows that ${\cal B}$ is the same for the 
$L_y=3$ triangular lattice strips with torus and Klein bottle topology, just as
was true of the analogous square lattice strips \cite{tk}, and in accord with 
the general discussion of Ref. \cite{bcc}.  For this case the generating
function has $deg_x({\cal N})=2$ and $deg_x({\cal D})=5$, so that, as was true
of the $L_y=3$ M\"obius and torus strips of the triangular lattice, all of 
the $\lambda_j$'s in ${\cal D}$ contribute to $P$.

For the sum of the coefficients we find, for the triangular $L_y=3$ strip with
torus boundary conditions 
\beq
C=P(K_3,q)=q(q-1)(q-2) \quad {\rm for} \quad G=tri(L_y=3) \quad {\rm with} 
\quad (PBC_y,PBC_x)
\label{csumtt}
\eeq
and for this strip with Klein bottle boundary conditions
\beq
C=0 \quad {\rm for} \quad G=tri(L_y=3) \quad {\rm with} \quad (PBC_y,TPBC_x) 
\ . 
\label{csumtk}
\eeq
These are the same respective values as those calculated in \cite{tk} for the 
corresponding $L_y=3$ strips of the square lattice with torus and Klein bottle
boundary conditions. 

\begin{figure}[p]
\centering
\leavevmode
\epsfxsize=4.0in
\begin{center}
\leavevmode
\epsffile{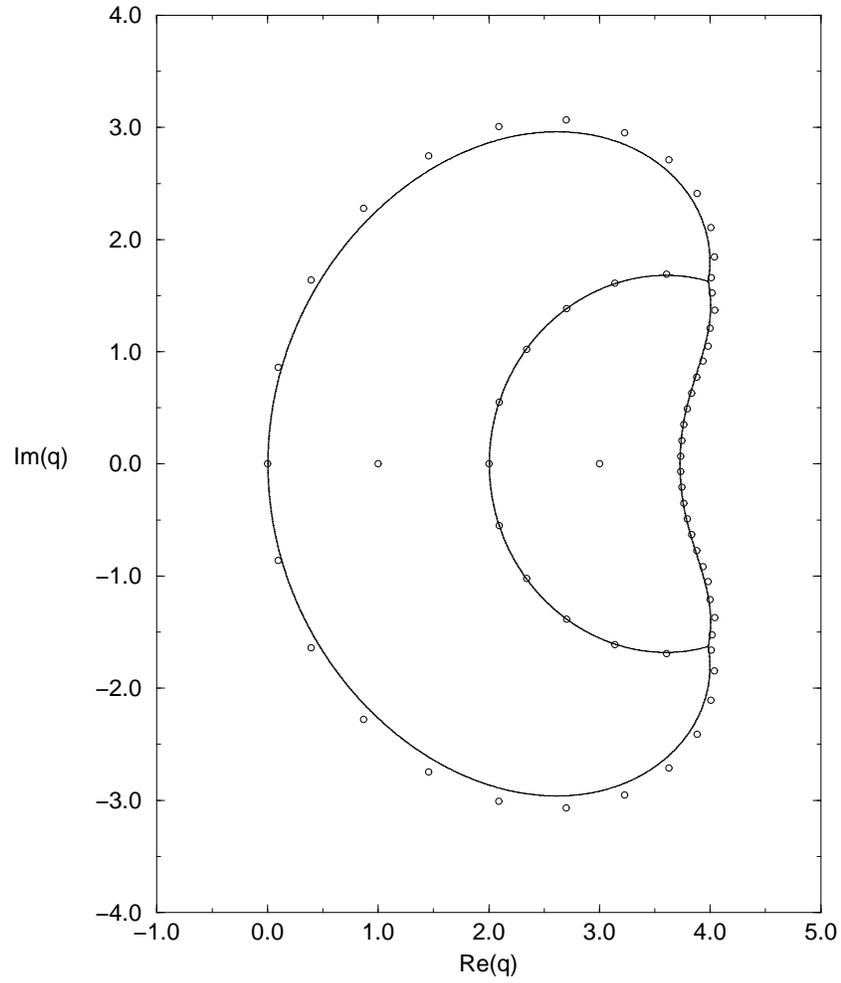}
\end{center}
\caption{\footnotesize{${\cal B}$ and chromatic zeros for the $L_y=3$,
$L_x=m=20$ (i.e., $n=60$) triangular strip graph with torus boundary
conditions.}}
\label{tritorzeros}
\end{figure}

Chromatic zeros for the $L_y=3$, $L_x=m=20$ torus graph of the triangular
lattice are shown in Fig. \ref{tritorzeros}; as before, with this value of
$m$, the chromatic zeros for the torus and Klein bottle boundary conditions are
generally similar and both sets of chromatic zeros lie close to the boundary
${\cal B}$ and indicate its position.  The locus ${\cal B}$ and the $W$
functions are the same for the torus and Klein bottle graph families.  We find
\beq
q_c \simeq 3.7240756... \quad {\rm for} \quad \{G\} = tri(L_y=3) \quad 
{\rm with} \quad (PBC_y,PBC_x) \ {\rm or} \ (PBC_y,TPBC_x) \ . 
\label{qcttorus}
\eeq
This value of $q_c$ is given by the real root of the degeneracy equation 
$q^3-9q^2+32q-46=0$. It is interesting that this is just 7 \% below the value
for the infinite 2D triangular lattice, $q_c(tri)=4$.  Thus, for the width 
$L_y=3$ strips where a comparison can be made, the 
use of both periodic transverse boundary conditions and periodic or twisted 
periodic longitudinal boundary conditions expedites the approach to the 2D
thermodynamic limit, in the sense that it yields a value of $q_c$ that is
substantially closer to the 2D value than was obtained for the same width strip
with free transverse boundary conditions in eq. (\ref{qctcycly3}).  This is
expected since for the torus or Klein bottle boundary conditions (a) the
resultant graphs are $\Delta$-regular\footnote{\footnotesize{A $\Delta$-regular
graph is a graph all of whose vertices have the same degree, $\Delta$, where
the degree of a vertex is defined as the number of edges connected to it.}}
with the degree (coordination number) $\Delta=6$ of the 2D triangular 
lattice, and (b) there are no boundaries to the surface on which the graphs are
embedded.  In contrast, for the boundary conditions of types (i)-(iv) (aside
from degenerate cases) the graphs are not $\Delta$-regular and they do have
boundaries.  

The locus ${\cal B}$ has support for $Re(q) \ge 0$ and
separates the $q$ plane into three regions.  The outermost one, region $R_1$,
extends to infinite $|q|$ and includes the intervals $q \ge q_c$ and $q \le 0$
on the real $q$ axis.  Region $R_2$ includes the real interval $2 \le q \le
q_c$ and extends upward and downward to the complex conjugate triple points on
${\cal B}$ at $q_t, q_t^* \simeq 4.0 \pm 1.7i$.  Region $R_3$ is the innermost
one and includes the real interval $0 \le q \le 2$.  The maximum value of
$|Im(q)|$ on the boundary between $R_1$ and $R_3$ is about 3.0, occurring at
$Re(q) \simeq 2.6$.  The maximum value of $|q|$ on ${\cal B}$ is roughly 4.6,
occurring at $q \simeq 3.9 \pm 2.5i$.  The boundary between $R_2$ and $R_3$
curves to the right as one increases $|Im(q)|$, extending from $q=2$ upward to
$q_t$ and downward to $q_t^*$.  As is evident in Fig. \ref{tritorzeros}, the
density of chromatic zeros is highest on the $R_1-R_2$ boundary near $q_c$.
The feature that ${\cal B}$ has support only for $Re(q) \ge 0$ is the same as
was found \cite{tk} for the $L_y=3$ strips of the square lattice with torus and
Klein bottle topology.

In region $R_1$, $\lambda_{tt,5}=\lambda_{tk,5}$ is the dominant 
$\lambda_j$, so 
\beq
W = (q^3-9q^2+29q-32)^{1/3} \ , \quad q \in R_1 \ . 
\label{wr1}
\eeq
The fact that this is the same as $W$ for the (PBC$_y$,FBC$_x$) case,
eq. (\ref{wsqly3pbc}), is a general result \cite{bcc}.  The importance of the
PBC$_y$ is evident from the fact that for the same width of three squares, the
strip with (FBC$_y$,FBC$_x$) yields a different $W$ \cite{strip}.

In region $R_2$ $\lambda_{tt,3}=\lambda_{tk,3}$ is dominant, so
\beq
|W| = |3q-14|^{1/3} \ , \quad q \in R_2
\label{wr2}
\eeq
(in regions other than $R_1$, only $|W|$ can be determined unambiguously 
\cite{w}). In region $R_3$, $\lambda_{tt,4}=\lambda_{tk,4}$ is dominant, so 
\beq
|W|=|2(q-4)^2|^{1/3} \ , \quad q \in R_3 \ . 
\label{wr3}
\eeq 
At all of the three points, $q=0,2$, and $q=q_c \simeq 3.72$ where
${\cal B}$ crosses the real $q$ axis, it does so vertically.  The present
results are in accord with the inference \cite{strip,wcy} that for a recursive
graph with regular lattice structure, a sufficient condition for
${\cal B}$ to separate the $q$ plane into two or more regions is that it
contains a global circuit, i.e. a path along a lattice direction whose length
goes to infinity as $n \to \infty$; here this is equivalent to PBC$_x$. 

Our calculations of the zero-temperature Potts antiferromagnet partition
functions (chromatic polynomials) and exponential of the entropy, $W$, for
$L_y=3$ strips of the triangular lattice with various boundary conditions
including free and periodic or reversed-orientation periodic 
thus elucidate the role that these boundary conditions and the
associated topologies play.
One particular feature of note is that the torus and Klein bottle graphs have 
interestingly different chromatic polynomials, with different $N_\lambda$, 
but the $W$ functions and hence the boundaries ${\cal B}$ are the same.

\section{Width $L_y=5,6$ Strips of the Triangular Lattice with $(PBC_y,FBC_x)$}

We first briefly review the cases $L_y=3,4$ \cite{strip2,w2d,bcc}.
The chromatic polynomial for the $L_y=3$ triangular $(t)$ strip with 
$n=3(m+2)$ vertices (following the labelling conventions in \cite{strip})
and $(PBC_y,FBC_x)$ boundary conditions, denoted $t3PF$ in 
the subscripts, has $N_\lambda=1$, 
\beq
\lambda_{t3PF}=q^3-9q^2+29q-32 \ , 
\label{lamtpy3}
\eeq
\beq
P(tri(3 \times m,PBC_y,FBC_x),q)=q(q-1)(q-2)(\lambda_{t3PF})^{m+1} \ , 
\label{ptpy3}
\eeq
and 
\beq
W(tri(L_y=3),PBC_y,FBC_x,q) = (q^3-9q^2+29q-32)^{1/3}
\label{wsqly3pbc}
\eeq
with ${\cal B}=\emptyset$.

For the $L_y=4$ strip with $(PBC_y,FBC_x)$ boundary conditions and 
$n=4(m+2)$ vertices, one had \cite{strip2,w2d} $N_\lambda=2$ and 
\beq
P(tri(4 \times m,PBC_y,FBC_x),q)=\sum_{j=1}^2 c_{t4PF,j}
(\lambda_{t4PF,j})^{m+1}
\label{ptri4pf}
\eeq
where
\beq
\lambda_{t4PF,j} = \frac{(q-3)}{2}\biggl [ T_{t4PF} \pm
(q-4)\sqrt{R_{t4PF}} \biggr ]
\label{lamtpy4}
\eeq
where 
\beq
T_{t4PF} = q^3-9q^2+33q-48
\label{tt4pf}
\eeq
\beq
R_{t4PF}=q^4-10q^3+43q^2-106q+129 \ . 
\label{rt4pf}
\eeq
In this case, as is evident in Fig. 4(a) of \cite{strip2}), ${\cal B}$ includes
both arcs and a closed oval invariant under complex conjugation which crossed
the real axis where the prefactor of the square root vanished, i.e. at
$q_c=4$, and at the real root of $T_{t4PF}$, namely $q \simeq 3.481406$.  There
are two regions, $R_1$ and $R_2$, the exterior and interior of the oval.  We
have 
\beq
W(tri, 4 \times \infty, PBC_y,q) = (\lambda_{t4PF,1})^{1/4} \ , \quad q \in R_1
\label{wt4pfr1}
\eeq
and
\beq
W(tri, 4 \times \infty, PBC_y,q) = (\lambda_{t4PF,2})^{1/4} \ , \quad q \in R_2
\label{wt4pfr2}
\eeq
with appropriate choices of branch cuts.

We have calculated new results for the width $L_y=5$ and $L_y=6$ strips with 
$(PBC_y,FBC_x)$ boundary conditions and $n=L_y(m+2)$ vertices.  For $L_y=5$ we 
find $N_\lambda=2$ and 
\beq
P(tri(5 \times m,PBC_y,FBC_x),q)=\sum_{j=1}^2 c_{t5PF,j}
(\lambda_{t5PF,j})^{m+1}
\label{ptri5pf}
\eeq
where
\beq
\lambda_{t5PF,j} = \frac{1}{2}\biggl [ T_{t5PF} \pm
S_{t5PF}\sqrt{R_{t5PF}} \biggr ]
\label{lamtpy5}
\eeq
where
\beq
T_{t5PF}=q^5-15q^4+98q^3-355q^2+711q-614
\label{tt5pf}
\eeq

\beq
R_{t5PF}=q^4-8q^3+26q^2-60q+85
\label{rt5pf}
\eeq
and
\beq
S_{t5PF}=q^3-11q^2+43q-58 \ . 
\label{st5pf}
\eeq
As before, the coefficients $c_{t5PF,j}$ can be computed using
eq. (\ref{chrompgsumlam}) in terms of the generating function.  This generating
function is given in the appendix.

Just as was true for the $L_y=4$ triangular lattice strip with $(PBC_y,FBC_x)$,
in the present $L_y=5$ case ${\cal B}$ includes both a pair of complex
conjugate arcs and an oval that crosses the real axis at two points. 
Analogously to the previous case, one of these points is the real zero of 
$T_{t5PF}$ at
\beq
q_\ell \simeq  3.207224
\label{qell}
\eeq
and the other is the real zero of the
prefactor $S_{t5PF}$ in front of the square root, at
\beq
q= 3.284775 = q_c  \quad {\rm for} \quad \{G\}=tri(5 \times \infty,
PBC_y,FBC_x) \ . 
\label{qctpy5}
\eeq
Taking the approximate center of the oval as the average of the left and right
crossings, $q_{center}=(q_\ell+q_c)$, one finds that
\beq
q_{center} \simeq 3.2460
\label{qcenter}
\eeq
near to the Tutte-Beraha number $B_7 \simeq 3.246980..$, where \cite{tutte,bkw}
\beq
B_r = 2 + 2\cos \Bigl ( \frac{2\pi}{r} \Bigr ) \ . 
\label{br}
\eeq 
Chromatic zeros near $B_7$ were also noted in \cite{baxter} for these
types of strips.  Note that $q_c$ decreases from 4 to the above value in
eq. (\ref{qctpy5}) as $L_y$ increases from 4 to 5.  Hence, in contrast to our
calculations of strips of the square and triangular lattices with
$(FBC_y,FBC_x)$, $(FBC_y,(T)PBC_x)$, and $(PBC_y,(T)PBC_x)$, where in all cases
considered, the respective $q_c$'s were nondecreasing functions of $L_y$, here
we find that this is not the case for boundary conditions of the type
$(PBC_y,FBC_x)$.

\begin{figure}[hbtp]
\centering
\leavevmode
\epsfxsize=2.5in
\begin{center}
\leavevmode
\epsffile{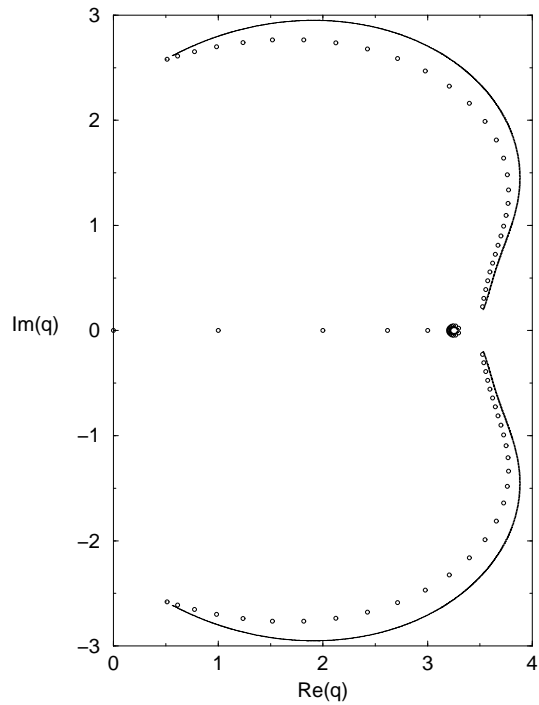}
\end{center}
\caption{\footnotesize{Locus ${\cal B}$ for the width $L_y=5$ strip (tube) of
the triangular lattice with $(PBC_y,FBC_x)$ boundary conditions.  Thus, the
cross sections of the tube form pentagons.  For comparison, chromatic zeros
calculated for the strip length $L_x=m+2=16$ (i.e., $n=80$ vertices) are shown.
}}
\label{tpy5}
\end{figure}

Chromatic zeros and ${\cal B}$ for this strip are shown in Fig. \ref{tpy5};
${\cal B}$ separates the $q$ plane into two regions, $R_1$, the exterior, 
and $R_2$, the interior, of the oval shown in the figure. 
Comparing the ovals for the $L_y=4$ strip (Fig. 4(a) of \cite{strip2}) and the
$L_y=5$ strip (Fig. \ref{tpy5} here), one sees that the oval shrinks in size.
We have 
\beq
W(tri, 5 \times \infty, PBC_y,q) = (\lambda_{t5PF,1})^{1/5} \ , \quad q \in R_1
\label{wt5pfr1}
\eeq
and 
\beq
W(tri, 5 \times \infty, PBC_y,q) = (\lambda_{t5PF,2})^{1/5} \ , \quad q \in R_2
\label{wt5pfr2}
\eeq
with appropriate choices of branch cuts. 

\vspace{6mm}

For the strip of with $L_y=6$ and $(PBC_y,FBC_x)$ we find $N_\lambda=5$ and
\beq
P(tri(6 \times m,PBC_y,FBC_x),q)=\sum_{j=1}^5 c_{t6PF,j}
(\lambda_{t6PF,j})^{m+1}
\label{ptri6pf}
\eeq
Here the $\lambda_{t6PF}$'s are the roots of a fifth-order equation, so we
cannot solve for them in terms of algebraic expressions as was possible for
previous strips of this type with $L_y=3,4,5$.  We give the generating function
for this strip in the appendix.  

\begin{figure}[hbtp]
\centering
\leavevmode
\epsfxsize=2.5in
\begin{center}
\leavevmode
\epsffile{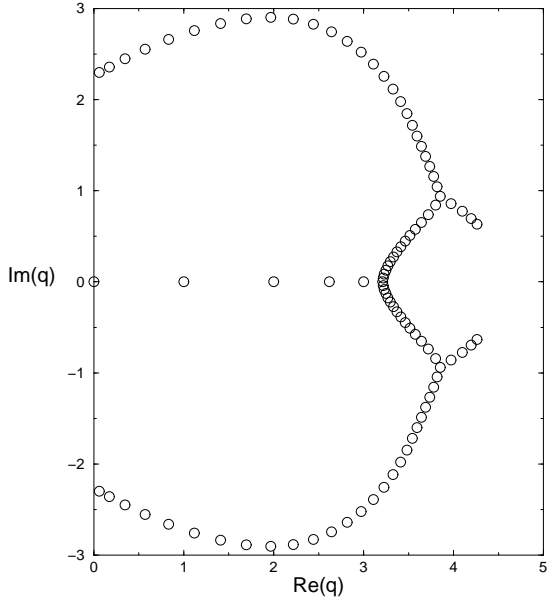}
\end{center}
\caption{\footnotesize{Locus ${\cal B}$ for the width $L_y=6$ strip (tube) of
the triangular lattice with $(PBC_y,FBC_x)$ boundary conditions.  Thus, the
cross sections of the tube form hexagons.  For comparison, chromatic zeros
calculated for the strip length $L_x=m+2=16$ (i.e., $n=96$ vertices) are shown.
}}
\label{tpy6}
\end{figure}

Chromatic zeros for this strip are shown in Fig. \ref{tpy6} for $L_x=m+2=16$, 
i.e., $n=96$; this length is sufficiently great that the chromatic zeros 
give an approximate indication of the location of the locus ${\cal B}$.  The
locus thereby inferred appears to consist of a single connected set of curves 
and crosses the real axis.  From our exact analytic expressions, we calculate 
the this crossing point to be 
\beq
q_c = 3.252419 \quad {\rm for} \quad \{G\}=tri(6 \times \infty,PBC_y,FBC_x) \ .
\label{qctpy6}
\eeq 
The morphology of chromatic zeros for this long $6 \times 16$ cylindrical
strip is similar to that found for a $8 \times 8$ patch with cylindrical
boundary conditions in \cite{baxter}.  In both cases, one can figuratively
think of how the inferred loci ${\cal B}$ can be modified to yield the locus
for the $L_y=\infty$ strip with $(PBC_y,FBC_x)$ cylindrical boundary
conditions: one (i) pulls the left-hand endpoints further to the left (into the
$Re(q) < 0$ half-plane) and around so that they meet in a cusp at the origin,
$q=0$; (ii) pulls the right-hand endpoints of the prongs over and around so
that they meet in a cusp at $q=4$; (iii) shifts the crossing curve slightly to
the right so that it crosses the real axis at $q \simeq 3.82$; and (iv) makes
minor shifts of the rest of the locus ${\cal B}$ so as to obtain ${\cal B}$ for
the cylindrical strip with $L_y=\infty$.  In this respect, just as was
discussed in \cite{w2d}, one sees that the periodic transverse boundary
conditions help to minimize finite-size effects.  Clearly, for a fixed $L_y$,
with $L_x \to \infty$, finite-size effects are reduced the most for $(PBC_y,
(T)PBC_x)$, next-most for $(PBC_y,FBC_x)$ and $(FBC_y,(T)PBC_x)$, and least for
$(FBC_y,FBC_x)$.  We find

\beq
W(tri,6 \times \infty,PBC_y,q)=(\lambda_{t6PF,max})^{1/6} \quad {\rm for} \ \
q \in R_1
\label{wt6pf}
\eeq
where $\lambda_{t6PF,max}$ denotes the root with the maximal magnitude. 

It is of interest to use these results to study the
approach of $W$ to the limit for the full infinite 2D triangular lattice,
extending the work of \cite{w2d}.  In Table \ref{tpf} we list
the various values of $W(tri(L_y \times \infty, PBC_y, FBC_x),q)$, denoted as
$W(L_y,q)$ to save space, together with the corresponding values of $W$ for the
full 2D triangular lattice, $W(tri,q)$, obtained via numerical evaluation of
an infinite product representation in \cite{baxter} as checked by series
expansions and rigorous bounds \cite{wn}, and the ratio $R_W(tri(L_y \times
\infty),PBC_y,FBC_x,q)$, where 
\beq
R_W(tri(L_y \times \infty),BC_y,BC_x,q) = \frac{W(tri(L_y \times \infty),
BC_y,BC_x,q)}{W(tri,q)} \ . 
\label{rw}
\eeq
This extends the previous study in \cite{w2d}.  Recall that for $q \ge 4$, 
$W(tri(L_y \times \infty),BC_y,BC_x,q)$ is independent of $BC_x$
\cite{w2d,bcc}.  Evidently, the approach of 
$W$ for the infinite-length finite-width triangular strips to the 2D 
thermodynamic limit as $L_y$ increases is quite rapid; for moderate values of 
$q$, say 6 or 7, the ratio $R_W(tri(5 \times\infty),PBC_y,FBC_x,q)$ is equal
to 1 to within approximately $10^{-3}$ or better, and the ratio 
$R_W(tri(6 \times\infty),PBC_y,FBC_x,q)$ is at least as close to 1.  As was 
proved in \cite{w2d}, the approach is non-monotonic for $PBC_y$ (and 
monotonic for $FBC_y$). 

\begin{table}
\caption{\footnotesize{Comparison of values of $W(tri(L_y),PBC_y,q)$ with
$W(tri,q)$ for $4 \le q \le 10$ and $BC_x=FBC_x$ or $(T)PBC_x$. For each value
of $q$, the quantities in the upper line are identified at the top and the
quantities in the lower line are the values of $R_W(tri(L_y),PBC_y,q)$.  The
$PBC_y$ is symbolized as $P_y$ in the table.}}
\begin{center}
\begin{tabular}{|c|c|c|c|c|c|}
\hline\hline
$q$ & $W(tri(3),P_y,q)$ & $W(tri(4),P_y,q)$ & $W(tri(5),P_y,q)$ & 
$W(tri(6),P_y,q)$ & $W(tri,q)$ \\
\hline\hline
4  & 1.58740   & 1.18921    & 1.39252  & 1.49603  &  1.46100   \\
   & 1.0865    & 0.8140     & 0.9531   & 1.0240   &  1         \\ \hline
5  & 2.35133   & 2.21336    & 2.26877  & 2.26894  &  2.26411   \\
   & 1.0385    & 0.9776     & 1.0021   & 1.0021   &  1         \\ \hline
6  & 3.23961   & 3.185055   & 3.20718  & 3.20399  &  3.20388   \\
   & 1.0112    & 0.9941     & 1.0010   & 1.0000   &  1         \\ \hline
7  & 4.17934   & 4.15965    & 4.16987  & 4.16805  &  4.16819   \\
   & 1.0027    & 0.99795    & 1.0004   & 1.0000   &  1         \\ \hline
8  & 5.14256   & 5.13936    & 5.14446  & 5.14348  &  5.14358   \\
   & 0.99980   & 0.9992     & 1.0002   & 1.0000   &  1         \\ \hline
9  & 6.11803   & 6.12324    & 6.12587  & 6.12533  &  6.12539   \\
   & 0.99880   & 0.99965    & 1.0001   & 1.0000   &  1         \\ \hline
10 & 7.10059   & 7.11027    & 7.11161  & 7.11131  &  7.11134   \\
   & 0.99849   & 0.99985    & 1.0000   & 1.0000   &  1       \\ \hline\hline
\end{tabular}
\end{center}
\label{tpf}
\end{table}

\section{Width $L_y=5$ Strips of the Triangular Lattices with 
$(FBC_y,FBC_x)$}

Previous studies have been published on strips of various lattices with
$(FBC_y,FBC_x)$ boundary conditions \cite{strip,strip2,hs}.  In the case of the
the square and triangular lattice, these went up to $L_y=4$ and involved
$\lambda_j$'s that were roots of a cubic and quartic equation, respectively.
Although one can calculate chromatic polynomials for wider strips, the analysis
is more cumbersome if the equations defining the $\lambda_j$'s are higher than
quartic, so that one cannot solve for these $\lambda_j$'s as analytic
closed-form algebraic expressions.  A study of these on wider open strips is in
\cite{ss}.  As an illustration of this sort of situation, here we present a
calculation of chromatic polynomials for the strip of the triangular lattice
with width $L_y=5$ and $(FBC_y,FBC_x)$.  For this width, the $\lambda_j$'s are
solutions of a degree-9 equation and hence cannot be solved for as algebraic
roots.  The method of calculation is again the iterated use of the
deletion-contraction theorem.  In \cite{strip}, a given strip $(G_s)_m$ was
constructed by $m$ successive additions of a subgraph $H$ to an endgraph $I$;
here, $I=H$, so that, following the notation of \cite{strip}, the total length
of the strip graph $(G_s)_m$ is $m+2$ vertices, or equivalently, $m+1$ edges in
the longitudinal direction.  The results are conveniently expressed in terms of
the coefficient functions in the generating function, as discussed above.  For
the width $L_y=5$ open strip of the triangular lattice we find $deg_z({\cal
D})= N_\lambda = 9$.  The coefficient functions $b_{t5FF,j}$ and $A_{t5FF,j}$
(cf. eqs. (\ref{d}) and (\ref{n})) are listed in the Appendix, where $FF$ is
short for $(FBC_y,FBC_x)$.

\begin{figure}[hbtp]
\centering
\leavevmode
\epsfxsize=2.5in
\begin{center}
\leavevmode
\epsffile{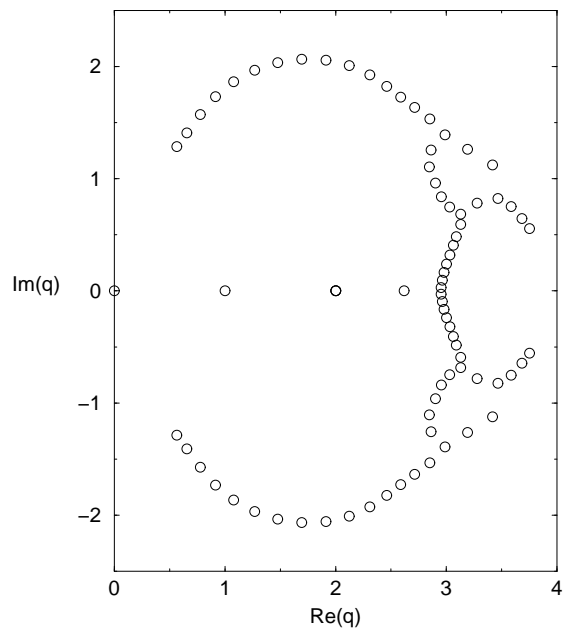}
\end{center}
\caption{\footnotesize{Chromatic zeros for the $L_y=5$ open strip of the
triangular lattice of length $m+2=16$ vertices (i.e. total number of vertices
$n=80$).}}
\label{ty5}
\end{figure}

In Fig. \ref{ty5} we show a plot of chromatic zeros for the open strip of the
triangular lattice with $L_y=5$ and length $m+2=16$ vertices, so that the strip
has $n=80$ vertices in all.  From an analysis of the degeneracy of leading
$\lambda_j$'s, we find the exact result 
\beq
q_c = 3 \quad {\rm for} \quad sq(5 \times \infty,FBC_y,FBC_x) \ . 
\label{qcsqff}
\eeq
This is in agreement with the chromatic zeros for the finite strip shown in 
Fig. \ref{ty5}.  Again, we can compare Fig. \ref{ty5} with the corresponding 
plots for $L_y=2$ and $L_y=3$ (Fig. 5(a,b) of \cite{strip}), and, just as was
true for the corresponding three open strips of the square lattice, this
comparison shows that as $L_y$ increases, the arcs forming ${\cal B}$ elongate 
and the arc endpoints nearest to the origin approach more closely to the 
origin.  We recall that for the $L_y=4$ open triangular strip, no $q_c$ is
defined since ${\cal B}$ does not cross the real axis.  

For $q > q_c$, we have, for the physical ground state degeneracy per site of
the $q$-state Potts antiferromagnet,
\beq
W(t(5 \times \infty,FBC_x, FBC_y), q) = (\lambda_{tri,j,max})^{1/5}
\label{wtri5}
\eeq 
where $\lambda_{sq,j,max}$ denotes the solution of eq. (\ref{xieq}) with
the coefficients (\ref{btri1})-(\ref{btri9}) that has the maximal magnitude in
region $R_1$.

As before, it is of interest to use this result to study the approach of $W$ to
the limit for the full infinite 2D triangular lattice, extending the work of
\cite{w2d} for this set of boundary conditions.  In Table \ref{tritable} we
list the various values of $W(tri(L_y \times \infty, FBC_y, BC_x),q)$, denoted
as $W(L_y,q)$ to save space, together with the corresponding values of $W$ for
the full 2D triangular lattice, $W(tri,q)$ and the ratio
$R_W(tri(L_y \times \infty),FBC_y,BC_x,q) = W(tri(L_y \times \infty),
FBC_y,BC_x,q)/W(tri,q)$.  Recall that the value of $W$ for $q \ge 4$ is
independent of $BC_x$ \cite{w2d}.  Again, the approach of $W$ for the
infinite-length finite-width triangular strips to the 2D thermodynamic limit as
$L_y$ increases is quite rapid.

\begin{table}
\caption{\footnotesize{Comparison of values of $W(tri(L_y \times \infty),q)$
for $FBC_y$ and $BC_x=FBC_x$ or $(T)PBC_x$ with $W(tri,q) \equiv W(tri(\infty
\times \infty),q)$ for $3 \le q \le 10$. For brevity of notation, we omit the
$(FBC_y,BC_x)$ in the notation.  For each value of $q$, the quantities in the
upper line are identified at the top and the quantities in the lower line are
the values of $R_W(tri(L_y),q)$.}}
\begin{center}
\begin{tabular}{|c|c|c|c|c|c|c|}
\hline\hline
$q$ & $W(tri(2),q)$ & $W(tri(3),q)$ & $W(tri(4),q)$ & $W(tri(5),q)$ & 
$W(tri(\infty),q)$
& \\ \hline\hline
4  &   2    & 1.77173   & 1.67619   & 1.62270   & 1.46100  \\
   & 1.369  & 1.213     & 1.147     & 1.111     & 1        \\ \hline
5  &   3    & 2.72998   & 2.60495   & 2.53251   &  2.26411 \\
   & 1.325  & 1.206     & 1.151     & 1.1185    & 1        \\ \hline
6  &   4    & 3.71457   & 3.579715  & 3.50112   &  3.20388 \\
   & 1.248  & 1.159     & 1.117     & 1.093     & 1        \\ \hline
7  &   5    & 4.70571   & 4.56515   & 4.48283   & 4.16819  \\
   & 1.200  & 1.129     & 1.095     & 1.075     & 1        \\ \hline
8  &   6    & 5.69974   & 5.55530   & 5.47040   & 5.14358  \\
   & 1.167  & 1.108     & 1.080     & 1.0635    & 1        \\ \hline
9  &   7    & 6.695395  & 6.54810   & 6.46129   & 6.12539  \\
   & 1.143  & 1.093     & 1.069     & 1.055     & 1        \\ \hline
10 &   8    & 7.69208   & 7.54259   & 7.45430   & 7.11134  \\
   & 1.125  & 1.082     & 1.061     & 1.048     & 1        \\ \hline\hline
\end{tabular}
\end{center}
\label{tritable}
\end{table}

\section{Comparative Discussion}

In this section we give a general discussion of the locus ${\cal B}$.  We have
found several interesting features: 

\begin{enumerate}

\item

For the strips of the triangular lattice of width $L_y=3$ and $L_y=4$ with
$(FBC_y,(T)PBC_x)$ and strips of width $L_y=3$ and $(PBC_y,(T)PBC_x)$ boundary
conditions studied here, we have shown that the locus ${\cal B}$ encloses
regions of the $q$ plane including certain intervals on the real axis and
passes through $q=0$ and 2 as well as other possible points, depending on the
family.  This extends the previous study of the $L_y=2$ strip of the triangular
lattice with $(FBC_y,(T)PBC_x)$ \cite{wcy}.  While the strips
with $(FBC_y,(T)PBC_x)$ have a locus ${\cal B}$ that passes through $q=3$, this
is not the case with at least the $L_y=3$ strip with $(PBC_y,(T)PBC_x)$.  For
the strips with $(FBC_y,(T)PBC_x)$ considered here we have shown that $q_c$ is
a nondecreasing function of $L_y$.  As one increases $L_y$, the $L_y=4$ cyclic
strip is the first one for which one can no longer solve for all of the
$\lambda_j$'s as algebraic roots, since some of the equations involved are
higher than quartic.

\item 

The crossing of ${\cal B}$ at the point $q=2$ for the (infinite-length limit
of) strips with global circuits nicely signals the property that the Ising
antiferromagnet has a frustrated zero-temperature critical point on these
strips.  This has been discussed in \cite{ta} in the context of exact solutions
for finite-temperature Potts model partition functions on the $L_y=2$ cyclic
and M\"obius triangular strips.  In contrast, this connection is not, in
general, present for strips with free longitudinal boundary conditions since
${\cal B}$ does not pass through $q=2$ (as recapitulated below).  The $q=3$
Potts antiferromagnet also has a zero-temperature critical point on these 
strips, and this is similarly manifested by the crossing of the singular locus
${\cal B}$ through the point $q=3$ for the cyclic and M\"obius strips with 
$(FBC_y,(T)BC_x)$, but not for the $L_y=3$ torus and Klein bottle strips with
$(PBC_y,(T)PBC_x)$. 

\item

An interesting feature of the cyclic strips of the triangular lattice (as well
as the square and kagom\'e lattices) is that for all of cases that have been
studied here and in \cite{w,wcy,pm}, there is a correlation between the
coefficient $c_{G_s,j}$ of the respective dominant $\lambda_{G_s,j}$'s in
regions that include intervals of the real axis.  Before, it was shown
\cite{bcc} that the $c_{G_s,j}$ of the dominant $\lambda_{G_s,j}$ in region
$R_1$ including the real intervals $q > q_c(\{G\})$ and $q < 0$ is $c^{(0)}=1$,
where the $c^{(d)}$ were given in eqs. (\ref{cd}), (\ref{cdzeros}). 
Here we extend this, observing that the $c_{G_s,j}$ that multiplies the
dominant $\lambda_{G_s,j}$ in the region containing the intervals $0 < q < 2$
is $c^{(1)}$.  For the cyclic strips of the triangular lattice that we have
studied, namely, $L_y=2,3,4$, the $c_{G_s,j}$ multiplying the dominant
$\lambda_{G_s,j}$ in the region containing the interval $2 < q < 3$ is
$c^{(2)}$.  For the $L_y=4$ strip of the triangular lattice, there is another
region containing the real interval $3 < q < q_c$, where $q_c$ for this strip
was given in (\ref{qctcycly4}), and we find that $c^{(3)}$ multiplies
the dominant $\lambda_{t4,j}$ containing this interval.  For the cyclic strips
of the square lattice, although the values of $q_c$ are different, a similar
correlation is observed; in particular, for the respective widths $L_y=3,4$, 
$c^{(2)}$ multiplies the $\lambda_{G_s,j}$ that is dominant in the region
containing the interval $2 < q < q_c$, where $q_c \simeq 2.34$ and 
$q_c \simeq 2.49$ for these two widths \cite{s4}. 

\item 

For the strip of the triangular lattice of width $L_y=5$ with cylindrical
$(PBC_y,FBC_x)$ boundary conditions, we find that ${\cal B}$ consists of arcs
together with a (closed) oval. However, ${\cal B}$ does not pass through $q=0,
2$, or 3.  This is qualitatively the same morphology that was found for the
corresponding strip of width $L_y=4$ (see Fig. 4 of \cite{strip2}).  For
comparison, the triangular strip with width $L_y=3$ and $(PBC_y,FBC_x)$ had
${\cal B}= \emptyset$.  The point $q_c=4$ for $L_y=4$ and $q_c \simeq 3.28$ for
$L_y=5$, which shows that for this family of strips, in cases where there is a
$q_c$ (there is none for $L_y=3$), it is not, in general, a nondecreasing
function of $L_y$.  This is somewhat reminiscent of the non-monotonicity of $W$
that we showed in the case of $(PBC_y,BC_x)$ strips, where $BC_x=FBC_x$ or
$(T)PBC_x$, in \cite{w2d}.  For the $L_y=6$ strip with $(PBC_y,FBC_x)$, we
infer from the chromatic zeros that ${\cal B}$ consists of a single component,
and from the analytic results we compute that $q_c \simeq 3.25$, which again
shows the nonmonotonicity of $q_c$ as a function of $L_y$. The morphology of
chromatic zeros for our $6 \times 16$ strip is similar to that found in
\cite{baxter} for $8 \times 8$ patch, both with cylindrical boundary
conditions.  For strips with torus or Klein bottle boundary conditions, $W$ and
${\cal B}$ have been calculated for only one $L_y$ value, namely, $L_y=3$ for
the square lattice in \cite{tk} and for the triangular lattice here.  For the
$L_y=3$ $(PBC_y,(T)PBC_x)$ square strip, $q_c$ is equal to the value 3 for the
full 2D square lattice, but for the $L_y=3$ $(PBC_y,(T)PBC_x)$ triangular
strip, our result (\ref{qcttorus}) shows that $q_c$ is less than the value of 4
for the 2D triangular lattice.  

\item 

We have included an illustrative result for a wider strip with $(FBC_y,FBC_x)$
boundary conditions, namely the strip with $L_y=5$.  As one increases $L_y$,
this is the first value at which one can no longer solve analytically for the
$\lambda_j$'s as algebraic roots.  The locus ${\cal B}$ is similar to the
respective loci that were found earlier in studies of open strips
\cite{strip,strip2,hs} in that it does not pass through $q=0$ or $q=2$ and does
not separate the $q$ plane into regions containing intervals of the real axis.
(It does pass through $q=3$, unlike the loci for the open strips with
$L_y=3,4$.) Our results confirm the trend that was observed in earlier work
\cite{strip,strip2}, namely that as $L_y$ increases, the arcs elongate and move
closer together, and the arc endpoints nearest to the origin move toward this
point.  In contrast to the strips with global circuits, these strips
do not manifest the property that the $q=2$ (Ising) and $q=3$ Potts
antiferromagnets have zero-temperature critical points since ${\cal B}$ does
not pass through these respective points.  The simplest example of this is the
Ising antiferromagnet on the infinite open line; in this case, although the
model has a well-known zero-temperature critical point, this is not evident in
the locus ${\cal B}$, which is the emptyset.

\item

For the strips with $(PBC_y,FBC_x)$ and $(FBC_y,FBC_x)$, while the endpoints of
the arcs on ${\cal B}$ that lie closest to the origin tend to move toward the
origin as $L_y$ increases, leading one to expect that in the limit $L_y \to
\infty$, the limiting locus ${\cal B}$ would pass through $q=0$, no such motion
toward $q=2$ is observed in the cases so far calculated.  This is in agreement
with the fact that the locus found in \cite{baxter} for the triangular lattice
constructed as the limit $L_x, L_y \to \infty$ with $(PBC_y,FBC_x)$ boundary
conditions passes through $q=0$ and 4 (and at $q \simeq 3.82$) but not through
$q=2$ or $q=3$.  Thus, assuming that our conjecture (\ref{bcrossq023}) is
correct, it follows that the locus ${\cal B}$ for the triangular lattice
depends on the boundary conditions used to define this lattice: if one
constructs as the limit $L_y \to \infty$ with $(PBC_y,FBC_x)$ (cylindrical)
boundary conditions, then ${\cal B}$ does not pass through $q=2$ or 3
\cite{baxter}, while if one constructs it as the limit $L_y \to \infty$ with
$(FBC_y,(T)PBC_x)$ (cyclic or M\"obius) boundary conditions, then, if the
conjecture is valid, ${\cal B}$ would pass through $q=2$ and 3 in the limit
just as it does for each of the values $L_y=2,3,4$ studied so far.  However,
since the value of $q_c$ pertains directly to a physical quantity, as the
minimal value of real $q$ above which $W(q)$ is analytic, one expects that
$q_c$ should be independent of the boundary conditions used to define the 2D
lattice.  All results obtained so far are consistent with this expectation.

\item

There have been a number of theorems proved concerning real chromatic zeros.
An elementary result is that no chromatic zeros can lie on the negative real
axis $q < 0$, since a chromatic polynomial has alternating coefficients.  It
has also been proved that there are no chromatic zeros in the intervals $0 < q
< 1$, and $1 < q < 32/27$ \cite{jackson}.  The bound of 32/27 in \cite{jackson}
has been shown to be sharp; i.e., for any $\epsilon > 0$, there exists a graph
with a chromatic zero at $q=32/27 + \epsilon$ \cite{thomassen}. See also 
\cite{brown,sokal} and references therein.  Based on our studies of strips of
the triangular (and square) lattices with the various boundary conditions
considered, we make the following observation: for such strips, we have not
found any chromatic zeros, except for the zero at $q=1$, in the interior of the
disk $|q-1|=1$.  This motivates the conjecture that for these strips, there are
no chromatic zeros with $|q-1| < 1$ except for the zero at $q=1$.  Assuming
that this conjecture is valid, the bound would be a sharp bound, since the 
circuit graph with $n$ vertices, $C_n$, has chromatic zeros lying precisely 
on the circle $|q-1|=1$ and at $q=1$ \cite{wc}. Further work is needed to prove
(or disprove) this conjecture. 

\end{enumerate} 

Some relevant features are summarized in Table \ref{proptable}.  The entries
for $L_y=\infty$ with $(PBC_y,FBC_x)$ are from \cite{baxter}. 

\begin{table}
\caption{\footnotesize{Comparative listing of properties of chromatic
polynomials $P$, ground state degeneracy functions $W$, and nonanalytic loci
${\cal B}$ for strip graphs $G_s$ of the triangular (tri) lattice and their
infinite-length limits.  New results in this work are marked with asterisks in
the first column. The properties apply for a given strip of type $G_s$ of size
$L_y \times L_x$; some apply for arbitrary $L_x$, such as $N_\lambda$, while
others apply for the infinite-length limit, such as the properties of the locus
${\cal B}$.  For the boundary conditions in the $y$ and $x$ directions ($BC_y$,
$BC_x$), F, P, and T denote free, periodic, and orientation-reversed (twisted)
periodic, and the notation (T)P means that the results apply for either
periodic or orientation-reversed periodic. The column denoted eqs. describes
the numbers and degrees of the algebraic equations giving the $\lambda_{G_s,j}$
in $P$; for example, $\{3(1),2(2),1(3)\}$ indicates that there are 3 linear
equations, 2 quadratic equations and one cubic equation.  The column denoted
BCR lists the points at which ${\cal B}$ crosses the real $q$ axis; the largest
of these is $q_c$ for the given family $G_s$. The notation ``none'' in this
column indicates that ${\cal B}$ does not cross the real $q$ axis.  Column
labelled ``SN'' refers to whether ${\cal B}$ has \underline{s}upport for
\underline{n}egative $Re(q)$, indicated as yes (y) or no (n).}}
\begin{center}
\begin{tabular}{|c|c|c|c|c|c|c|c|}
\hline\hline $G_s$ & $L_y$ & $BC_y$ & $BC_x$ & $N_\lambda$ & eqs. & BCR & SN
\\ \hline\hline
tri & 2 & F & (T)P & 4  & \{2(1),1(2)\} & 3, \ 2, \ 0 & n  \\ \hline
**tri & 3 & F & (T)P & 10 & \{3(1),2(2),1(3)\} & 3, \ 2, \ 0 & n \\ \hline
**tri & 4 & F &   P & 26 & \{1(1),2(4),1(8),1(9)\} & 3.23, \ 3, \ 2, \ 0 & y \\
\hline\hline
tri & 3 & P   & F  & 1  & \{1(1)\}  & none  & $-$           \\ \hline
tri & 4 & P   & F  & 2  & \{1(2)\}  & 4, \ 3.48  & n        \\ \hline
**tri& 5 & P  & F  & 2  & \{1(2)\}  & 3.28, \ 3.21 & n     \\ \hline
**tri& 6 & P  & F  & 5  & \{1(5)\}  & 3.25        & n      \\ \hline 
tri&$\infty$& P& F  & $-$& $-$       & 4, \ 3.82, 0 & y     \\ \hline\hline
**tri & 3 & P    & P & 11 & \{5(1),3(2)\} & 3.72, \ 2, \ 0 & n \\ \hline
**tri & 3 & P   & TP & 5  & \{5(1)\}      & 3.72, \ 2, \ 0 & n \\ \hline\hline
tri & 2 & F   & F & 1  & \{1(1)\} & none   & $-$          \\ \hline
tri & 3 & F   & F & 2  & \{1(2)\} & 2.57   & n            \\ \hline
tri & 4 & F   & F & 4  & \{1(4)\} & none   & n            \\ \hline
**tri& 5 & F   & F & 9  & \{1(9)\} & 3     & n            \\ \hline\hline
\end{tabular}
\end{center}
\label{proptable}
\end{table}

\section{Conclusions}

In this paper we have presented exact calculations of the zero-temperature
$q$-state Potts antiferromagnet partition functions (equivalently, chromatic
polynomials $P$), on strips of the triangular lattice of width $L_y=3$ and with
boundary conditions of four types: (a) $(FBC_y,PBC_x)=$ cyclic, (b)
$(FBC_y,TPBC_x)=$ M\"obius, (c) $(PBC_y,PBC_x)=$ toroidal, and (d)
$(PBC_y,TPBC_x)=$ Klein bottle, where $F$, $P$, and $TP$ denote free, periodic,
and twisted periodic. In the infinite-length limits of these strips, exact
results were given for the ground state degeneracy (exponential of the ground
state entropy), $W$, and its analytic structure in the complex $q$ plane, in
particular, the nonanalytic locus ${\cal B}$, was discussed.  Exact
calculations of $P$ and $W$ and studies of ${\cal B}$ were also presented for
wider strips, including (e) cyclic, $L_y=4$, (f) $(PBC_y,FBC_x)=$ cylindrical,
$L_y=5,6$, and an illustrative $(FBC_y,FBC_x)=$ open case with $L_y=5$.  A
comparative analysis of these results was included.  An interesting result of
our calculations of $W$ on infinite-length strips with different widths and
transverse boundary conditions is the observation that for the cases studied,
${\cal B}$ passes through $q=2$ (as well as $q=0$) for strips with periodic or
twisted periodic longitudinal boundary conditions but does not for strips with
free longitudinal boundary conditions.  Hence, in particular, if one uses
periodic or twisted periodic longitudinal boundary conditions, the locus ${\cal
B}$ nicely signals the existence of the zero-temperature critical point of the
Ising antiferromagnet on these infinite-length, finite-width strips of the
triangular lattice.  Numerical values of $W$ were given for infinite-length
strips of various widths and were shown to approach values for the 2D lattice
rapidly.  Some conjectures for the behavior of ${\cal B}$ for arbitrarily wide
strips, and for a region in the $q$ plane free of chromatic zeros, were stated.
These exact calculations of the $T=0$ Potts antiferromagnet partition function
and ground state degeneracy on strips of the triangular lattice give valuable
analytic knowledge of properties of Potts antiferromagnets.

\vspace{10mm}

Note added: The original version of this paper, submitted in early Oct. 1999,
contained calculations on $L_y=3$ strips of the triangular lattice.  In
response to a request by a referee to perform calculations for wider strips, we
have added the results on the $L_y=4$ strip with $(FBC_y,PBC_x)$, the
$L_y=5,6$ strips with $(PBC_y,FBC_x)$, and the $L_y=5$ strip with 
$(FBC_y,FBC_x)$.

\vspace{10mm}

Acknowledgment: The research of R. S. was supported in part by the U. S. NSF
grant PHY-97-22101. 

\section{Appendix}

\subsection{Generating Function for $L_y=3$ M\"obius Strip of the Triangular
Lattice} 

As noted in the text, it is convenient to leave the coefficients $c_{t3Mb,j}$, 
$j=6,7,8$ in the general form (\ref{ctmj}).  For the evaluation of 
these coefficients, we list here the generating function for this strip.  
We have $d_{\cal N}=8$ and $d_{\cal D}=10$, and 
\beq
{\cal D}(tri(L_y=3),FBC_y,TPBC_x,q,x) = \prod_{j=1}^{10}
(1-\lambda_{t3,j}(q)x)
\label{lambdaformtmb}
\eeq
where the $\lambda_{t3,j}$ were given in eq. (\ref{ptcyclic}).  
Since several of the $\lambda_{t3,j}$'s are algebraic, it is useful to 
display the denominator in an explicitly polynomial form:
\beqs
& & {\cal D}(tri(L_y=3),FBC_y,TPBC_x,q,x)=(1+x)\Bigl [ 1-(q-2)x \Bigr ] \times
\cr\cr
& & \Bigl [1-(2q-7)x+(q-2)(q-3)x^2\Bigr ]
\Bigl [ 1+(q-2)(q-3)x \Bigr ]F_{t3q3}F_{t3q2}
\label{dtly3mb}
\eeqs
where
\beq
F_{t3q3}=\prod_{j=6,7,8}(1-\lambda_{t3,j}x)=1+b_{t3,1}x+b_{t3,2}x^2+b_{t3,3}x^3
\label{ft3q3}
\eeq
with $b_{t3,j}$, $j=1,2,3$ given in eqs. (\ref{bt31})-(\ref{bt33}), and 
\beq
F_{t3q2}=\prod_{j=9,10}(1-\lambda_{t,j}x)=
 1-(q^3-7q^2+18q-17)x+(q-2)^3(q-3)x^2 \ . 
\label{ft3q2}
\eeq
For the numerator, extracting a common factor via the definition of the reduced
coefficients $\bar A_{t3Mb,j}$, 
\beq
A_{t3Mb,j} \equiv q(q-1)(q-2)(q-3)\bar A_{t3Mb,j} \ , 
\label{areduced}
\eeq
we have
\beq
\bar A_{t3Mb,0}=q^2-6q+10
\label{atm0}
\eeq
\beq
\bar A_{t3Mb,1}=(q-3)(q^3-12q^2+45q-55)
\label{atm1}
\eeq
\beq
\bar A_{t3Mb,2}=q^6-20q^5+170q^4-779q^3+2016q^2-2779q+1588
\label{atm2}
\eeq
\beq
\bar A_{t3Mb,3}=-(q-3)(5q^6-82q^5+574q^4-2185q^3+4745q^2-5536q+2691)
\label{atm3}
\eeq
\beq
\bar A_{t3Mb,4}=(q-2)(q-3)(9q^6-141q^5+930q^4-3303q^3+6651q^2-7176q+3224)
\label{atm4}
\eeq
\beq
\bar A_{t3Mb,5}=-(q-2)^2(q-3)^3(6q^4-55q^3+186q^2-277q+152)
\label{atm5}
\eeq
\beq
\bar A_{t3Mb,6}=-(q-2)^5(q-3)(q^4-12q^3+59q^2-138q+125)
\label{atm6}
\eeq
\beq
\bar A_{t3Mb,7}=(q-2)^6(q-3)^3 (3q^2-14q+17)
\label{atm7}
\eeq
\beq
\bar A_{t3Mb,8}=-(q-2)^9(q-3)^3 \ . 
\label{atm8}
\eeq

It should also be noted that there is a significant difference between the
strips of the triangular lattice studied here and the analogous strips of the
square lattice \cite{wcy,pm,tk}.  In general, a strip of the square lattice of
width $L_y$ and length $L_x$ with any of the boundary conditions
$(FBC_y,PBC_x)$ (cyclic), $(FBC_y,TPBC_x)$ (M\"obius), $(PBC_y,PBC_x)$ (torus),
or $(PBC_y,TPBC_x)$ (Klein bottle) is invariant under a translation by one edge
or vertex in the longitudinal direction.  However, in the case of the analogous
strip of the triangular lattice, this is only true of the cases with cyclic and
torus boundary conditions; the strips with M\"obius and Klein bottle boundary
conditions have a ``seam'' along with the orientation of the triangles
reverses.  This is discussed further in the appendix.  Thus, if one proceeds in
a longitudinal direction along the triangular-lattice M\"obius strip, starting
in a manner such that the triangles are formed by edges connecting the upper
left and lower right vertices of squares (relative to one's direction of
motion), then when one crosses this seam, the triangles will be formed by edges
connecting the upper right and lower left vertices of the squares on the strip.
Related to this, there are differences in the degrees of various vertices on
these strips (where the degree of a vertex is defined as the number of edges
that connect to this vertex).  If one avoids the lowest few values of $L_x$
where the strips degenerate, then, in general, (i) for the cyclic triangular
strip of width $L_y$ and length $L_x$, the $(L_y-2)L_x$ internal vertices have
degree 6 while the $2L_x$ vertices on the upper and lower sides have degree 4;
(ii) for the same strip as in (i) but with M\"obius instead of cyclic
longitudinal boundary conditions, the $(L_y-2)L_x$ internal vertices have
degree 6, the $2(L_y-1)$ vertices on the upper and lower sides except for those
on the seam have degree 4, and, on the seam, the external vertices have degrees
5 and 3; (iii) for the same strip as in (i) but with torus or Klein bottle
boundary conditions, all of the vertices have degree 6.

\subsection{Equations for the Terms in the Chromatic Polynomial for the 
Cyclic $L_y=4$ Strip of the Triangular Lattice}

Four of the $\lambda_{t4,j}$, which we label as $j=2,3,4,5$, are the same as 
for the open $L_y=4$ strip of the triangular lattice.  These are solutions to 
the quartic equation
\beq
\xi^4+b_{t4,1,1}\xi^3+b_{t4,1,2}\xi^2+b_{t4,1,3}\xi+b_{t4,1,4}=0
\label{t4eq1}
\eeq
where the coefficients were given as $b_{t(4),j} \equiv b_{t4,1,j}$, 
$j=1,..,4$ in eqs. (B.15)-(B.18) of \cite{strip}

The terms $\lambda_{t4,j}$, $j=6,7,8,9$ are solutions to the quartic equation 
\beq
\xi^4+b_{t4,2,1}\xi^3+b_{t4,2,2}\xi^2+b_{t4,2,3}\xi+b_{t4,2,4}=0
\label{t4eq2}
\eeq
where
\beq
b_{t4,2,1}=4q-13
\label{bt421}
\eeq
\beq
b_{t4,2,2}=2(3q^2-18q+26)
\label{bt422}
\eeq
\beq
b_{t4,2,3}=(q-2)(4q^2-25q+38)
\label{bt423}
\eeq
and
\beq
b_{t4,2,4}=(q-2)^2(q-3)^2 \ . 
\label{bt424}
\eeq

Another set of $\lambda_j$'s for $10 \le j \le 17$ are roots of an equation 
of degree 8,
\beq
\xi^8 + \sum_{k=1}^8 b_{t4,3,k}\xi^{8-k}=0
\label{t4eq3}
\eeq
where
\beq
b_{t4,3,1}=2(-3q^2+19q-31)
\label{bt431}
\eeq
\beq
b_{t4,3,2}=15q^4-186q^3+867q^2-1794q+1385
\label{bt432}
\eeq
\beq
b_{t4,3,3}=-20q^6+366q^5-2784q^4+11248q^3-25425q^2+30452q-15080
\label{bt433}
\eeq
\beq
b_{t4,3,4}=(q-2)(15q^7-334q^6+3174q^5-16676q^4+52294q^3-97852q^2+101138q-44528)
\label{bt434}
\eeq
\beq
b_{t4,3,5}=-(q-2)^2(q-3)^2(6q^6-126q^5+1076q^4-4804q^3+11861q^2-15378q+8185)
\label{bt435}
\eeq
\beq
b_{t4,3,6}=(q-2)^4(q-3)^3(q^5-25q^4+216q^3-868q^2+1670q-1246)
\label{bt436}
\eeq
\beq
b_{t4,3,7}=(q-2)^6(q-3)^4(2q^3-22q^2+80q-97)
\label{bt437}
\eeq
\beq
b_{t4,3,8}=(q-2)^8(q-3)^6 \ . 
\label{bt438}
\eeq

A final set of $\lambda_{t4,j}$, $18 \le j \le 26$, are solutions to the 
equation of degree 9
\beq
\xi^9 + \sum_{k=1}^9 b_{t4,4,k}\xi^{9-k}=0
\label{t4eq4}
\eeq
where
\beq
b_{t4,4,1}=2(q-3)(2q^2-12q+21)
\label{bt441}
\eeq
\beq
b_{t4,4,2}=(q-3)(6q^5-90q^4+558q^3-1772q^2+2865q-1875)
\label{bt442}
\eeq
\beqs
& & b_{t4,4,3}=4q^9-111q^8+1380q^7-10071q^6+47476q^5-149742q^4+315652q^3
\cr\cr
& & -428385q^2+339300q-119368
\label{bt443}
\eeqs
\beqs
& & b_{t4,4,4}=(q-2)(q-3)^2(q^9-34q^8+491q^7-4032q^6+20961q^5-71954q^4 \cr\cr
& & +163654q^3-238278q^2+201722q-75672)
\label{bt444}
\eeqs
\beqs
& & b_{t4,4,5}=-(q-2)^2(q-3)^3(3q^9-81q^8+983q^7-7029q^6+32609q^5-101701q^4 
\cr\cr
& & +213036q^3-288702q^2+229385q-81299)
\label{bt445}
\eeqs
\beqs
& & b_{t4,4,6}=(q-2)^4(q-3)^4(3q^8-70q^7+728q^6-4404q^5+16929q^4-42286q^3 
\cr\cr
& & +66933q^2-61296q+24830)
\label{bt446}
\eeqs
\beq
b_{t4,4,7}=-(q-2)^6(q-3)^6(q^6-20q^5+166q^4-734q^3+1833q^2-2462q+1393)
\label{bt447}
\eeq
\beq
b_{t4,4,8}=-(q-2)^8(q-3)^7(2q^4-21q^3+87q^2-165q+119)
\label{bt448}
\eeq
\beq
b_{t4,4,9}=-(q-2)^{12}(q-3)^8 \ . 
\label{bt449}
\eeq

\subsection{Generating Function for the $L_y=5,6$ Strips of the Triangular
Lattice with $(PBC_y,FBC_x)$}

For the $L_y=5$ strip we calculate a generating function of the form
(\ref{gammagen}) with $d_{\cal D}=2$, $d_{\cal N}=1$ and, in the notation of
eqs. (\ref{n}) and (\ref{d}), 
\beq 
b_{t5PF,1}=-q^5+15q^4-98q^3+355q^2-711q+614
\label{bt5pf1}
\eeq

\beq
b_{t5PF,2}=(q-3)(3q^7-66q^6+619q^5-3205q^4+9877q^3-18065q^2+18078q-7588)
\label{bt5pf2}
\eeq

\beq
A_{t5PF,0}=q(q-1)(q-2)(q-3)(q^6-14q^5+85q^4-290q^3+599q^2-723q+398)
\label{at5pf0}
\eeq

\beqs
& & A_{t5PF,1}=-q(q-1)(q-2)(q-3)(q^2-2q+2)(3q^7-66q^6+619q^5-3205q^4 \cr\cr
& & +9877q^3-18065q^2+18078q-7588) \ . 
\label{at5pf1}
\eeqs

For the $L_y=6$ strip we calculate a generating function of the form 
(\ref{gammagen}) with $d_{\cal D}=5$, $d_{\cal N}=4$ and 
\beq
b_{t6PF,1}=-q^6+18q^5-145q^4+680q^3-1980q^2+3379q-2586
\label{bt6pf1}
\eeq

\beqs
& & b_{t6PF,2} = 4q^{10}-128q^9+1868q^8-16352q^7+94977q^6-382031q^5+1076317q^4 
\cr\cr & & -2093899q^3+2686606q^2-2047842q+702080
\label{bt6pf2}
\eeqs

\beqs
& & b_{t6PF,3} = -2(q-3)^2(q^{12}-39q^{11}+695q^{10}-7493q^9+54509q^8
-282283q^7+1068575q^6 \cr\cr 
& & -2982861q^5+6098756q^4-8908956q^3+8820488q^2-5306146q+1462992)
\label{bt6pf3}
\eeqs

\beqs
& & b_{t6PF,4} = -2(q-2)(q-3)^5(2q^{11}-62q^{10}+882q^9-7601q^8+44105q^7 
-181018q^6 \cr\cr 
& & +536536q^5-1149015q^4+1742334q^3-1779827q^2+1099188q-309188)
\label{bt6pf4}
\eeqs

\beqs
& & b_{t6PF,5} = 4(q-2)^2(q-3)^8(q^2-5q+5)(2q^7-36q^6+277q^5-1179q^4+2990q^3
\cr\cr & & -4505q^2+3728q-1310)
\label{bt6pf5}
\eeqs

\beqs
& & A_{t6PF,0} = q(q-1)(q-2)(q^9-21q^8+199q^7-1121q^6+4159q^5-10623q^4 \cr\cr
& & +18887q^3-22824q^2+17177q-6143)
\label{at6pf0}
\eeqs

\beqs
& & A_{t6PF,1} = -q(q-1)(q-2)(4q^{13}-140q^{12}+2266q^{11}-22416q^{10}+
150973q^9
\cr\cr & & -730186q^8+2607252q^7-6958852q^6+13899608q^5-20584349q^4 
\cr\cr & & +22103679q^3-16461349q^2+7723994q-1748140) 
\label{at6pf1}
\eeqs

\beqs
& & A_{t6PF,2} = 2q(q-1)(q-2)(q-3)^2(q^{15}-42q^{14}+816q^{13}-9734q^{12}
+79793q^{11}
\cr\cr & & -476549q^{10}+2144264q^9-7409966q^8+19852299q^7-41297346q^6
+66301130q^5 \cr\cr 
& & -80939629q^4+73099740q^3-46448750q^2+18742947q-3655548)
\label{at6pf2}
\eeqs

\beqs
& & A_{t6PF,3} = 2q(q-1)(q-2)(q-3)^5(2q^{15}-72q^{14}+1212q^{13}-12651q^{12}
+91556q^{11}
\cr\cr & & -486599q^{10}+1962326q^9-6116898q^8+14870220q^7-28223745q^6
+41554711q^5 \cr\cr & & -46735427q^4+39026466q^3-22975509q^2+8586616q-1545752)
\label{at6pf3}
\eeqs

\beqs
& & A_{t6PF,4}= -4q(q-1)(q-2)^2(q-3)^8(q^2-5q+5)(q^4-5q^3+10q^2-10q+5) \cr\cr
& & \times (2q^7-36q^6+277q^5-1179q^4+2990q^3-4505q^2+3728q-1310)
\label{at6pf4}
\eeqs

\subsection{Generating Function for $L_y=5$ Open Strip of the Triangular
Lattice}

For this strip we calculate a generating function of the form (\ref{gammagen})
with $d_{\cal D}=9$, $d_{\cal N}=8$ and, in the notation of eqs. (\ref{n})
and (\ref{d}),
\beq
b_{t5FF,1}=-(q-3)(q^4-10q^3+46q^2-113q+120)
\label{btri1}
\eeq

\beqs
& &
b_{t5FF,2}=6q^8-131q^7+1280q^6-7328q^5+26930q^4-65081q^3+100888q^2-91462q
+36965
\cr\cr
& &
\label{btri2}
\eeqs

\beqs
& & b_{t5FF,3}=-(q-2)(q-3)(15q^9-378q^8+4289q^7-28788q^6+126096q^5-374139q^4
\cr\cr
& & +752541q^3-989867q^2+772611q-272483)
\label{btri3}
\eeqs

\beqs
& & b_{t5FF,4}=(q-2)(q-3)^2(20q^{11}-607q^{10}+8429q^9-70702q^8+398115q^7
\cr\cr
& & -1580547q^6+4515585q^5-9285872q^4+13471537q^3-13131321q^2 \cr\cr
& & +7738560q-2087938)
\label{btri4}
\eeqs

\beqs
& & b_{t5FF,5}=-(q-2)^2(q-3)^3(15q^{12}-502q^{11}+7729q^{10}-72397q^9
\cr\cr
& & +459566q^8-2083176q^7+6915864q^6-16947196q^5+30430188q^4 \cr\cr
& & -39053679q^3+34008163q^2-18041392q+4408580)
\label{btri5}
\eeqs

\beqs
& & b_{t5FF,6}=(q-2)^3(q-3)^5(6q^{12}-203q^{11}+3149q^{10}-29630q^9
+188440q^8 -853749q^7 \cr\cr
& & +2826657q^6-6893466q^5+12293272q^4-15636918q^3+13466958q^2 \cr\cr
& & -7049578q+1695556)
\label{btri6}
\eeqs

\beqs
& & b_{t5FF,7}=-(q-2)^5(q-3)^7(q^{11}-34q^{10}+516q^9-4647q^8+27718q^7
-115308q^6 \cr\cr
& & +342008q^5-724072q^4+1072964q^3-1060043q^2+628196q-169014)
\label{btri7}
\eeqs

\beqs
& & b_{t5FF,8}=-(q-2)^7(q-3)^9(q^9-23q^8+241q^7-1505q^6+6145q^5-16929q^4
\cr\cr
& & +31319q^3-37359q^2+25972q-7987)
\label{btri8}
\eeqs

\beqs
& & b_{t5FF,9}=(q-2)^{12}(q-3)^{11}(q^3-8q^2+21q-17)
\label{btri9}
\eeqs

We use the definition $A_{t5FF,j}=q(q-1)(q-2) \bar A_{t5FF,j}$ and have 
\beq
\bar A_{t5FF,0}=(q-2)^7
\label{atri0}
\eeq

\beqs
& & \bar A_{t5FF,1}=-(q-2)^2(6q^8-113q^7+926q^6-4304q^5+12381q^4-22504q^3
\cr\cr
& & +25133q^2-15663q+4121)
\label{atri1}
\eeqs

\beqs
& & \bar A_{t5FF,2}=(q-2)^2(q-3)(15q^{10}-363q^9+3936q^8-25141q^7+104572q^6
\cr\cr
& & -295342q^5+572184q^4-748554q^3+630336q^2-306798q+65006)
\label{atri2}
\eeqs

\beqs
& &
\bar A_{t5FF,3}=-(q-2)^2(q-3)^2(20q^{12}-587q^{11}+7867q^{10}-63587q^9
+344801q^8 \cr\cr
& & -1319670q^7+3650087q^6-7338694q^5+10623070q^4-10770336q^3+7237158q^2
\cr\cr
& & -2882235q+511741)
\label{atri3}
\eeqs

\beqs
& &
\bar A_{t5FF,4}=(q-2)^3(q-3)^3(15q^{13}-487q^{12}+7266q^{11}-65902q^{10}
+405008q^9 \cr\cr
& & -1779077q^8+5739974q^7-13753341q^6+24433426q^5-31724357q^4+29191727q^3
\cr\cr
& & -17973781q^2+6613085q-1093258)
\label{atri4}
\eeqs

\beqs
& &
\bar A_{t5FF,5}=-(q-2)^4(q-3)^4(6q^{14}-215q^{13}+3554q^{12}-35907q^{11}
+247594q^{10}
\cr\cr
& & -1231849q^9+4556733q^8-12718548q^7+26883798q^6-42760478q^5+50287310q^4
\cr\cr
& & -42307379q^3+24007733q^2-8200007q+1267630)
\label{atri5}
\eeqs

\beqs
& &
\bar A_{t5FF,6}=(q-2)^6(q-3)^7(q^{12}-33q^{11}+485q^{10}-4225q^9+24378q^8
-98266q^7 \cr\cr
& & +283651q^6-589907q^5+875195q^4-900611q^3+607987q^2-240848q+42203) \cr\cr
& &
\label{atri6}
\eeqs

\beqs
& &
\bar A_{t5FF,7}=(q-2)^{10}(q-3)^9(q^8-18q^7+145q^6-674q^5+1941q^4-3474q^3
\cr\cr
& & +3701q^2-2120q+499)
\label{atri7}
\eeqs

\beq
\bar A_{t5FF,8}=-(q-1)^3(q-2)^{11}(q-3)^{11}(q^3-8q^2+21q-17) \ . 
\label{atri8}
\eeq

\subsection{$L_y=2$ Cyclic and M\"obius Strips of the Triangular Lattice 
with Odd $N_t$} 

\begin{figure}[h]
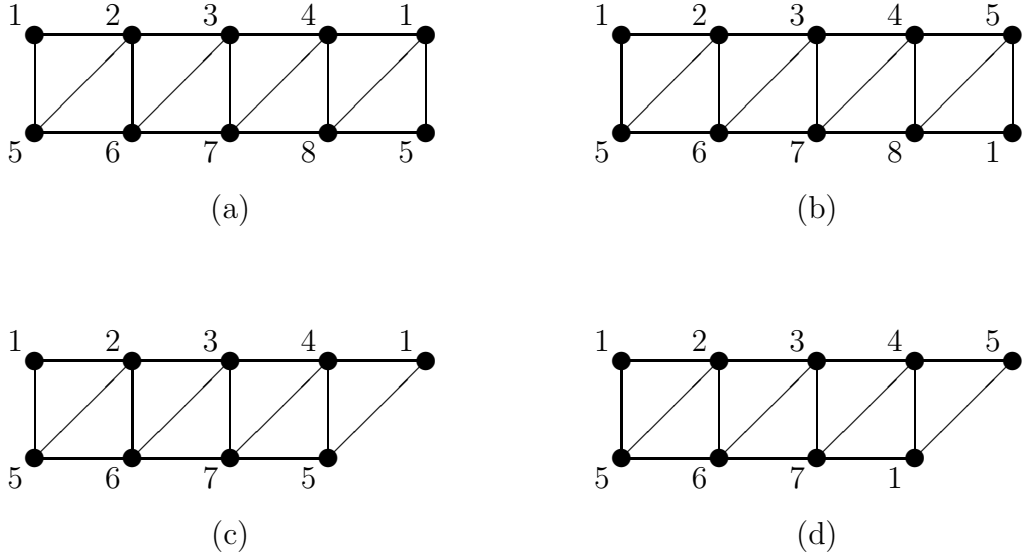

\caption{\footnotesize{Illustrative strip graphs of the triangular lattice with
width $L_y=2$: (a) cyclic, $L_x=4$, even $N_t=2L_x$; (b)
M\"obius, $L_x=4$, even $N_t=2L_x$; (c) cyclic, odd $N_t=7$;
(d) M\"obius, odd $N_t=7$.}}
\label{figseam}
\end{figure}

In this part of the appendix we shall report some new results for $L_y=2$
cyclic and M\"obius strips of the triangular lattice with an odd number $N_t$
of triangles and compare these with the case of even $N_t=L_xL_y$.
Some illustrative strip graphs are shown in Fig. \ref{figseam}.  

\vspace{10mm}

\unitlength 1.3mm

\begin{picture}(100,10)
\multiput(0,0)(10,0){5}{\circle*{2}}
\multiput(0,10)(10,0){5}{\circle*{2}}
\multiput(0,0)(10,0){5}{\line(0,1){10}}
\multiput(0,0)(0,10){2}{\line(1,0){40}}
\multiput(0,0)(10,0){4}{\line(1,1){10}}
\put(-2,-2){\makebox(0,0){5}}
\put(8,-2){\makebox(0,0){6}}
\put(18,-2){\makebox(0,0){7}}
\put(28,-2){\makebox(0,0){8}}
\put(38,-2){\makebox(0,0){5}}
\put(-2,12){\makebox(0,0){1}}
\put(8,12){\makebox(0,0){2}}
\put(18,12){\makebox(0,0){3}}
\put(28,12){\makebox(0,0){4}}
\put(38,12){\makebox(0,0){1}}
\put(20,-8){\makebox(0,0){(a)}}

\multiput(60,0)(10,0){5}{\circle*{2}}
\multiput(60,10)(10,0){5}{\circle*{2}}
\multiput(60,0)(10,0){5}{\line(0,1){10}}
\multiput(60,0)(0,10){2}{\line(1,0){40}}
\multiput(60,0)(10,0){4}{\line(1,1){10}}
\put(58,-2){\makebox(0,0){5}}
\put(68,-2){\makebox(0,0){6}}
\put(78,-2){\makebox(0,0){7}}
\put(88,-2){\makebox(0,0){8}}
\put(98,-2){\makebox(0,0){1}}
\put(58,12){\makebox(0,0){1}}
\put(68,12){\makebox(0,0){2}}
\put(78,12){\makebox(0,0){3}}
\put(88,12){\makebox(0,0){4}}
\put(98,12){\makebox(0,0){5}}
\put(80,-8){\makebox(0,0){(b)}}
\end{picture}
\vspace*{3cm}

\begin{picture}(100,10)
\multiput(0,0)(10,0){4}{\circle*{2}}
\multiput(0,10)(10,0){5}{\circle*{2}}
\multiput(0,0)(10,0){4}{\line(0,1){10}}
\put(0,0){\line(1,0){30}}
\put(0,10){\line(1,0){40}}
\multiput(0,0)(10,0){4}{\line(1,1){10}}
\put(-2,-2){\makebox(0,0){5}}
\put(8,-2){\makebox(0,0){6}}
\put(18,-2){\makebox(0,0){7}}
\put(28,-2){\makebox(0,0){5}}
\put(-2,12){\makebox(0,0){1}}
\put(8,12){\makebox(0,0){2}}
\put(18,12){\makebox(0,0){3}}
\put(28,12){\makebox(0,0){4}}
\put(38,12){\makebox(0,0){1}}
\put(20,-8){\makebox(0,0){(c)}}

\multiput(60,0)(10,0){4}{\circle*{2}}
\multiput(60,10)(10,0){5}{\circle*{2}}
\multiput(60,0)(10,0){4}{\line(0,1){10}}
\put(60,0){\line(1,0){30}}
\put(60,10){\line(1,0){40}}
\multiput(60,0)(10,0){4}{\line(1,1){10}}
\put(58,-2){\makebox(0,0){5}}
\put(68,-2){\makebox(0,0){6}}
\put(78,-2){\makebox(0,0){7}}
\put(88,-2){\makebox(0,0){1}}
\put(58,12){\makebox(0,0){1}}
\put(68,12){\makebox(0,0){2}}
\put(78,12){\makebox(0,0){3}}
\put(88,12){\makebox(0,0){4}}
\put(98,12){\makebox(0,0){5}}
\put(80,-8){\makebox(0,0){(d)}}
\end{picture}

\vspace{10mm}

For the $L_y=2$ cyclic strip of the triangular lattice with even $N_t=2L_x$ 
we calculated \cite{wcy} (see also \cite{sands,matmeth}) 
\beq
P(tri(L_y=2,cyc.),N_t=2m,q) = (q^2-3q+1) + [(q-2)^2]^m + (q-1)\Bigl [
(\lambda_{t2,3})^m + (\lambda_{t2,4})^m \Bigr ] 
\label{ptrinteven}
\eeq
where 
\beq
\lambda_{t2,(3,4)} = \frac{1}{2}\biggl [ 5-2q \pm \sqrt{9-4q} \ \biggr ] \ .
\label{lamtly234}
\eeq
For the corresponding $L_y=2$ M\"obius strip of the triangular lattice with 
even $N_t=2L_x$, we calculated \cite{wcy}
\beq
P(tri(L_y=2,Mb.),N_t=2m,q) = -1 + [(q-2)^2]^m -
(q-1)(q-3)\frac{\Bigl [ (\lambda_{t2,3})^m - (\lambda_{t2,4})^m \Bigr ]}{
\lambda_{t2,3}-\lambda_{t2,4}} \ . 
\label{ptrimbnteven}
\eeq
Note that the M\"obius strip has a seam.  The function $W$ and the boundary 
${\cal B}$ were given in \cite{wcy}; ${\cal B}$ separates the $q$ plane into 
three regions and crosses the real axis at $q=0,2$, and at $q_c=3$. 

We proceed to our new results.  For the $L_y=2$ cyclic strip
of the triangular lattice containing an odd number $N_t=2m+1$ of triangles, 
we calculate 
\beqs
& & P(tri(L_y=2,cyc.),N_t=2m+1,q) = -(q^2-3q+1) + (q-2)[(q-2)^2]^m + \cr\cr
& & \frac{1}{2}(q-1)(q-3)\Biggl [\Bigl ( (\lambda_{t2,3})^m+ (\lambda_{t2,4})^m
\Bigr ) + \frac{\Bigl ( (\lambda_{t2,3})^m - (\lambda_{t2,4})^m \Bigr )}{
\lambda_{t2,3}-\lambda_{t2,4}} \Biggr ] \ . 
\label{ptrintodd}
\eeqs
This strip has a seam. 

For the $L_y=2$ M\"obius strip of the triangular lattice with an odd number 
$N_t=2m+1$ of triangles (which does not have a seam) we obtain 
\beqs
& & P(tri(L_y=2,Mb.),N_t=2m+1,q) = 1 +  (q-2)[(q-2)^2]^m + \cr\cr
& & \frac{1}{2}(1-q)\Biggl [ \Bigl ( (\lambda_{t2,3})^m+ (\lambda_{t2,4})^m
\Bigr ) + (9-4q)\frac{\Bigl ( (\lambda_{t2,3})^m - (\lambda_{t2,4})^m \Bigr )}
{\lambda_{t2,3}-\lambda_{t2,4}} \Biggr ] \ . 
\label{ptrimbntodd}
\eeqs
Note that the sum of the coefficients $C=0$ for the chromatic polynomials for
both of the $L_y=2$ strips with odd $N_t$, eqs. (\ref{ptrintodd}) and 
(\ref{ptrimbntodd}).  For comparison, for the $L_y=2$ even-$N_t$ strips, 
$C=q(q-1)$ for the cyclic case, eq. (\ref{ptrinteven}) and $C=0$ for the
M\"obius case, eq. (\ref{ptrimbnteven}). 

Equivalently, one may write these results in terms of generating functions, 
with the definition analogous to (\ref{gamma}):
\beq
\Gamma(tri(L_y=2,BC_x),N_t=even,q,x) = \sum_{m=2}^{\infty}
P(tri(L_y=2,BC_x),N_t=2m,q)x^{m-2} \ . 
\label{gammatrintevengen}
\eeq
and 
\beq
\Gamma(tri(L_y=2,BC_x),N_t=odd,q,x) = \sum_{m=2}^{\infty}
P(tri(L_y=2,BC_x),N_t=2m+1,q)x^{m-2}
\label{gammatrintodd}
\eeq
where $BC_x=PBC_x$ or $TPBC_x$.  The generating functions for all of these
strips have the same denominator, 
\beqs
& & {\cal D}(tri(L_y=2,cyc.))=\prod_{j=1}^4 (1-\lambda_{t2,j}x) \cr\cr
& & = (1-x)\Bigl [ 1-(q-2)^2x \Bigr ] \Bigl [1-(5-2q)x+(q-2)^2x^2 \Bigr ] \ . 
\label{dtri}
\eeqs
The numerators are easily worked out from the results that we have given for
the chromatic polynomials and the denominator ${\cal D}$; for example, 
\beq
\Gamma(tri(L_y=2,cyc.),N_t=even,q,x) = 
\frac{ q(q-1)(q-2)\Bigl [ q-3+(q-2)x-(q-2)^3x^2 \Bigr ]}
{{\cal D}(tri(L_y=2,cyc.))}
\label{gammatrinteven}
\eeq
\beq
\Gamma(tri(L_y=2,cyc.),N_t=odd,q,x) = \frac{q(q-1)(q-2)(q-3)
\bigl [ q-3+(q-2)^2x \bigr ]}{{\cal D}(tri(L_y=2,cyc.))} \ . 
\label{gammatri}
\eeq
and so forth for the M\"obius strips. 

We remark on some general features of these results.  The chromatic polynomials
for both even and odd $N_t$ and both cyclic and M\"obius strips have the same 
four $\lambda_j$'s, and hence the same $W$ functions and boundary ${\cal B}$ 
(given in \cite{wcy}).  For all cases, $q_c=3$.  These properties are in
agreement with the general discussion in \cite{wcy,pm,bcc} on the effects of 
boundary conditions on $W$ and ${\cal B}$.

\vfill
\eject
\end{document}